%% file: main.tex
\documentclass{aastex631}

\shorttitle{Predicting Solar Flares using SMARPs and SHARPs}
\shortauthors{Sun et al.}

\graphicspath{{./}{figures/}}

\usepackage{amsmath} 
\usepackage[section]{placeins} 
\usepackage{booktabs}
\usepackage{multirow}

\usepackage{rotating}
\usepackage{makecell, multirow}
\renewcommand\theadfont{\normalsize}

\def \TP{\textrm{TP}}
\def \TN{\textrm{TN}}
\def \FP{\textrm{FP}}
\def \FN{\textrm{FN}}
\def \Ptrue{\textrm{P}} 
\def \Ntrue{\textrm{N}} 
\def \Ppred{\textrm{P}'} 
\def \Npred{\textrm{N}'} 
\def \Total{\textrm{N} + \textrm{P}} 
\def \F1{\textrm{F}_1}
\def \Fbeta{\textrm{F}_\Fbeta}
\def \ACC{\textrm{ACC}}
\def \TSS{\textrm{TSS}}
\def \HSS{\textrm{HSS}}
\def \AUC{\textrm{AUC}}
\def \BS{\textrm{BS}}
\def \BSS{\textrm{BSS}}

\def \fusedsharp{\texttt{FUSED\_SHARP}} 
\def \sharponly{\texttt{SHARP\_ONLY}}
\def \fusedsmarp{\texttt{FUSED\_SMARP}}
\def \smarponly{\texttt{SMARP\_ONLY}}

\begin{document}

\title{Predicting Solar Flares Using CNN and LSTM on Two Solar Cycles of Active Region Data}

\correspondingauthor{Zeyu Sun}
\email{zeyusun@umich.edu}

\author[0000-0003-3903-928X]{Zeyu Sun}
\affiliation{Department of Electrical Engineering and Computer Science, University of Michigan, Ann Arbor, MI 48105, USA}

\author[0000-0002-5662-9604]{Monica G. Bobra}
\affiliation{W.W. Hansen Experimental Physics Laboratory, Stanford University, Stanford, CA 94305, USA}

\author[0000-0002-8963-7432]{Xiantong Wang}
\affiliation{Department of Climate and Space Sciences and Engineering, University of Michigan, Ann Arbor, MI 48109, USA}

\author[0000-0002-6287-4710]{Yu Wang}
\affiliation{Department of Statistics, University of Michigan
Ann Arbor, MI 48109, USA}

\author{Hu Sun}
\affiliation{Department of Statistics, University of Michigan
Ann Arbor, MI 48109, USA}

\author[0000-0001-9360-4951]{Tamas Gombosi}
\affiliation{Department of Climate and Space Sciences and Engineering, University of Michigan, Ann Arbor, MI 48109, USA}

\author[0000-0002-9516-8134]{Yang Chen}
\affiliation{Department of Statistics, University of Michigan
Ann Arbor, MI 48109, USA}

\author[0000-0002-2531-9670]{Alfred Hero}
\affiliation{Department of Electrical Engineering and Computer Science, University of Michigan, Ann Arbor, MI 48105, USA}
\affiliation{Department of Statistics, University of Michigan
Ann Arbor, MI 48109, USA}

\begin{abstract}

We consider the flare prediction problem that distinguishes flare-imminent active regions that produce an M- or X-class flare in the future 24 hours, from quiet active regions that do not produce any flare within $\pm 24$ hours. Using line-of-sight magnetograms and parameters of active regions in two data products covering Solar Cycle 23 and 24, we train and evaluate two deep learning algorithms---CNN and LSTM---and their stacking ensembles. The decisions of CNN are explained using visual attribution methods. We have the following three main findings.
(1) LSTM trained on data from two solar cycles achieves significantly higher True Skill Scores (TSS) than that trained on data from a single solar cycle with a confidence level of at least 0.95.
(2) On data from Solar Cycle 23, a stacking ensemble that combines predictions from LSTM and CNN using the TSS criterion achieves significantly higher TSS than the ``select-best" strategy with a confidence level of at least 0.95.
(3) A visual attribution method called Integrated Gradients is able to attribute the CNN's predictions of flares to the emerging magnetic flux in the active region. It also reveals a limitation of CNN as a flare prediction method using line-of-sight magnetograms: it treats the polarity artifact of line-of-sight magnetograms as positive evidence of flares.


\end{abstract}

\section{Introduction} \label{sec:intro}

Solar flares are abrupt electromagnetic explosions occurring in magnetically active regions on the solar surface. Intense solar flares are frequently followed by coronal mass ejections and solar energetic particles, which may disturb or disable satellites, terrestrial communication systems, and power grids. Predicting such strong flares from solar observations is therefore of particular significance and has been one of the primary tasks in space weather research.

Flare prediction can be posed as a classification problem, asking for a binary decision on whether the sun will produce a flare above some level in a future time window. Since strong solar flares mostly occur in active regions, it is common to first produce predictions for each active region on the solar disk.
In this paper, we consider a ``strong-vs-quiet" flare prediction problem, distinguishing active regions that will produce an M- or X- class flare in the future 24 hours, from those that stay flare quiescent within 24 hours before---and after---the forecast issuance time.

Over the past decade, a great amount of flare prediction studies have been conducted on a data product named Space-Weather HMI Active Region Patches \citep[SHARPs,][]{bobra2014helioseismic}. The SHARP database is derived from full-disk observations of the Helioseismic and Magnetic Imager \citep[HMI,][]{schou12} aboard the {\it Solar Dynamics Observatory} (SDO), containing maps and summary parameters of automatically tracked active regions from May 2010 to the present day, covering much of Solar Cycle 24. Despite the fact that SHARP is one of the most recent and highest quality datasets of its kind, it only contains a limited number of strong events, as Solar Cycle 24 is the weakest solar cycle in a century. Recently, a new data product, Space-Weather MDI Active Region Patches \citep[SMARPs,][]{bobra2021smarps}, was developed as an effort to extend backward the SHARP database to include active region observations in Solar Cycle 23, a much stronger solar cycle with significantly more flaring events. In fact, Solar Cycle 23 is the longest solar cycle (147 months) in the past 150 years\footnote{Source: \url{https://ntrs.nasa.gov/api/citations/20130013068/downloads/20130013068.pdf}}. The SMARP database was derived from the Michelson Doppler Imager \citep[MDI,][]{scherrer95} aboard the {\it Solar and Heliospheric Observatory} (SoHO), which observed the sun from 1996 to 2010. Compared to its successor HMI, MDI's measurement of the solar surface magnetic field is only restricted to the line-of-sight component, with lower spatial resolution, lower signal-to-noise ratio, and shorter cadence. As such,
SMARP does not contain as much information as SHARP, and its data quality is not as high.
Nonetheless, SMARP's coverage of a stronger solar cycle and its partial compatibility with SHARP make it a valuable data product to use with SHARP, especially for statistical studies in which a large sample size or a long time span is desired.

Many  machine  learning  methods  for  flare  prediction  have  been  proposed  in  recent  years.
They roughly fall into three categories in terms of how flare pertinent features are extracted from data. The first category uses \emph{explicit} parameterization of observational data that are considered relevant to flare production, e.g., SHARP parameters that characterize the photospheric magnetic field. Much of the effort in data-driven flare forecasting has been made in this category, exploring a wide range of machine learning algorithms including discriminant analysis \citep{leka2003photospheric}, regularized linear regression \citep{jonas2018flare}, support vector machine \citep{yuan2010automated, bobra2015solar, nishizuka2017solar, florios2018forecasting}, k-nearest neighbors \citep{nishizuka2017solar}, extremely random trees \citep{nishizuka2017solar}, random forests \citep{liu2017predicting, florios2018forecasting}, multi-layer perceptrons (MLP) \citep{florios2018forecasting}, residual networks \citep{nishizuka2018deep, nishizuka2020reliable}, long short-term memory (LSTM) networks \citep{chen2019identifying, liu2019predicting}, etc. The second category learns features from images using fixed transformations, e.g., random filters \citep{jonas2018flare}, Gabor filters \citep{jonas2018flare}, wavelet transforms \citep{hada2016deep}. The third category, only popularized more recently, \emph{implicitly} learns flare indicative signatures directly from active region magnetic field maps. This category features mainly convolutional neural networks (CNNs) \citep{huang2018deep, li2020predicting}. Note that the three categories are not mutually exclusive. For example, methods in the second category typically also depend on explicitly constructed features \citep[e.g.][]{jonas2018flare} as the information within transformation coefficients is often limited.
In this study, two representative deep learning methods, LSTM and CNN, are considered. LSTM uses times series of active region summary parameters derived from line-of-sight magnetograms, whereas CNN uses static point-in-time
magnetograms.


With so many machine learning algorithms developed for flare forecasting, one might expect an improved performance by combining different methods. This expectation seems even more reasonable if component methods in the combination use different data to provide complementary information. This is the idea behind ensemble learning, a learning paradigm that capitalizes on different models to achieve better performance than is achievable by any of the models alone. During the past few decades, the rapidly-evolving field of ensemble learning has achieved great success in many areas, which has attracted the attention of the space weather community \citep{murray2018importance}. In this paper, using CNN and LSTM models independently trained on active region magnetograms and parameter sequences, respectively, we consider a particular type of ensemble method called stacking \citep{wolpert1992stacked}.
Ideas similar to our stacking ensemble method have previously appeared in solar flare forecasting, most notably \citet{guerra2015ensemble,guerra2020ensemble}. In their work, full-disk probabilistic forecasts are linearly combined with weights selected by maximizing a potentially non-convex performance metric (e.g., the True Skill Score, the Heidke Skill Score).
In contrast, our stacking ensemble linearly combines two state-of-the-art machine learning classifiers (LSTM and CNN). To select the combination weights, we consider a convex cross-entropy loss function in addition to other performance metrics.

Deep learning models are often considered to be ``black-box" due to lack of interpretability. Recently the machine learning community has developed empirical tools that aim to better interpret the decisions of the deep neural network (DNN). Among these tools is the class of attribution methods \citep[e.g.][]{springenberg2015striving,selvaraju2017grad,shrikumar2017learning,sundararajan2017axiomatic} that attribute a score to gauge the contribution of each input feature for a given input sample. Attribution methods such as the occlusion method and Grad-CAM have been previously used to interpret CNNs in flare prediction applications \citep{bhattacharjee2020supervised,yi2021visual}. In this work, we evaluate additional attribution methods (Deconvolution, Guided Backpropagation, DeepLIFT, and Integrated Gradients) on the interpretation of CNNs trained to predict flares. In particular, we show that Integrated Gradient attribution maps, which have the same resolution as the input image, lead to insights on the important magnetic features that inform the CNN's decisions on flare prediction. 


The contributions of this paper 
are as follows: 
\begin{enumerate}
    \item We demonstrate the value  of combining SMARP and SHARP to improve flare prediction performance.
    \item We compare the flare prediction performance of LSTM and CNN on an equal footing, i.e., on the same temporally evolving active region dataset.
    \item We demonstrate that stacking the LSTM and CNN can significantly improve flare class prediction in certain settings. 
    \item We provide visual explanations of the CNN predictor using visual attribution methods including Deconvolution, Guided Backpropagation, Integrated Gradients, DeepLIFT, and Grad-CAM. We demonstrate the potential of these methods in identifying flare indicative signatures, interpreting CNN's decisions, revealing model limitations, and suggesting model modifications.
\end{enumerate}

The rest of the paper is organized as follows. Section \ref{sec:data} introduces 
the data sources and how they are processed into machine-learning-ready datasets. Section \ref{sec:methodology} describes the flare prediction methods, stacking ensemble, and visual attribution methods. Section \ref{sec:results} presents and compares the flare prediction performance on the datasets. Section \ref{sec:discussion} concludes the paper by presenting the lessons learned from the experiments.

\section{Data}
\label{sec:data}



\subsection{Data sources}


Observational data of active regions of Solar Cycle 23 and 24 are extracted from the SMARP and the SHARP data product, respectively. Both SMARP and SHARP contain automatically-tracked active region cutouts of full-disk line-of-sight magnetograms, referred to as Track Active Region Patches (TARPs) and HMI Active Region Patches (HARPs), respectively. They also contain summary parameters that characterize physical properties of active regions. We consider definitive SMARP and SHARP records in Cylindrical Equal-Area (CEA) coordinates hosted in Joint Science Operations Center\footnote{See \url{http://jsoc.stanford.edu}.}. We query SMARP records from 1996 April 23 to 2010 October 28 and SHARP records from 2010 May 1 to 2020 December 1, both at a cadence of 96 minutes. Only good quality SMARP and SHARP records within $\pm 70^\circ$ of the central meridian matching at least one NOAA active region are considered.
For active region summary parameters, we use four metadata keywords that are common to SMARP and SHARP, i.e., \texttt{USFLUXL}, \texttt{MEANGBL}, \texttt{R\_VALUE}, and \texttt{AREA}. Definitions of these summary parameters are listed in Table \ref{tab:keywords}. For images, we use photospheric line-of-sight magnetic field maps, or magnetograms, from the two data products. 

\begin{table}
    \centering
    \caption{Active region summary parameters used in this study. The line-of-sight magnetic field is denoted by $B_{L}$.}
    \begin{tabular}{p{0.07\textwidth}p{0.24\textwidth}p{0.28\textwidth}p{0.2\textwidth}p{0.1\textwidth}}
        \toprule
        Keyword & Description & Pixels & Formula & Unit \\
        \midrule
        \texttt{USFLUXL} & Total line-of-sight unsigned flux & Pixels in the TARP/HARP region & $\sum|B_{L}|dA$ & Maxwell\\
        \texttt{MEANGBL} & Mean gradient of the line-of-sight field & Pixels in the TARP/HARP region & $\frac{1}{N}\sum \sqrt{\left(\frac{\partial B_{L}}{\partial x}\right)^2 + \left(\frac{\partial B_{L}}{\partial y}\right)^2}$ & Gauss/pixel \\
        \texttt{R\_VALUE} & $R$, or a measure of the unsigned flux near polarity inversion lines \citep{schrijver2007r} & Pixels near polarity inversion lines & $\log\left(\sum|B_{L}|dA\right)$ & Maxwell \\
        \texttt{AREA} & De-projected area of patch on sphere in micro-hemisphere & Pixels in the TARP/HARP region & $\sum dA$ & mH \\
        \bottomrule
    \end{tabular}
    \label{tab:keywords}
\end{table}

Observational data samples are labeled using the GOES catalog of X-ray solar flare events. Based on the peak magnitude of 1--8 \AA{} soft X-ray flux measured by \emph{Geostationary Operational Environmental Satellites} (GOES), solar flare events are classified into five increasingly intense classes: A, B, C, M, and X, sometimes appended with a number that indicates a finer scale classification. M- and X- class flares are referred to as strong flares throughout the paper.
We only consider GOES solar flare events with at least one associated NOAA active region that can be used to cross-reference the \texttt{NOAA\_ARS} keyword in SHARP (or SMARP) databases to associate the flare with a HARP (or TARP).
The GOES event records are queried using the Sunpy package \citep{sunpy_community2020} from the beginning of 1996 to the end of 2020, covering the period of the SMARP and SHARP observations used in this paper.  Of note, although the GOES catalog is widely considered as the ``go-to" record database in solar flare forecasting, it is not error-free. There are cases in which flares, even the strong ones, are not assigned to any active region \citep{leka2019comparison}. Furthermore, small-sized flares could be
buried under 
the background radiation, a phenomenon frequently observed for A- and B-class flares, especially after a strong flare occurs.
Moreover, there are 61 event records annotated with an unknown GOES event class, most of them in the year 1996. These 61 event records are excluded in this study.

\subsection{Data fusion}

The challenge of combining the disparate SHARP and SMARP data is mitigated by the fact that there is a short overlapping time period over which they were jointly collected (May 1 to October 28 of 2010). We used this common time period to evaluate the dissimilarities between the SHARP/SMARP data and to develop methods for data alignment. As explained below, our analysis of the data over the common time period led us to adopt a simple method for fusing the SHARP and SMARP data: (1) we downsampled the SHARP magnetograms to match the resolution of the SMARP magnetograms; (2) we separately transformed the SHARP and SMARP summary parameters by Z-score (translation-scale) standardization.

\begin{figure}
    \centering
    \gridline{\fig{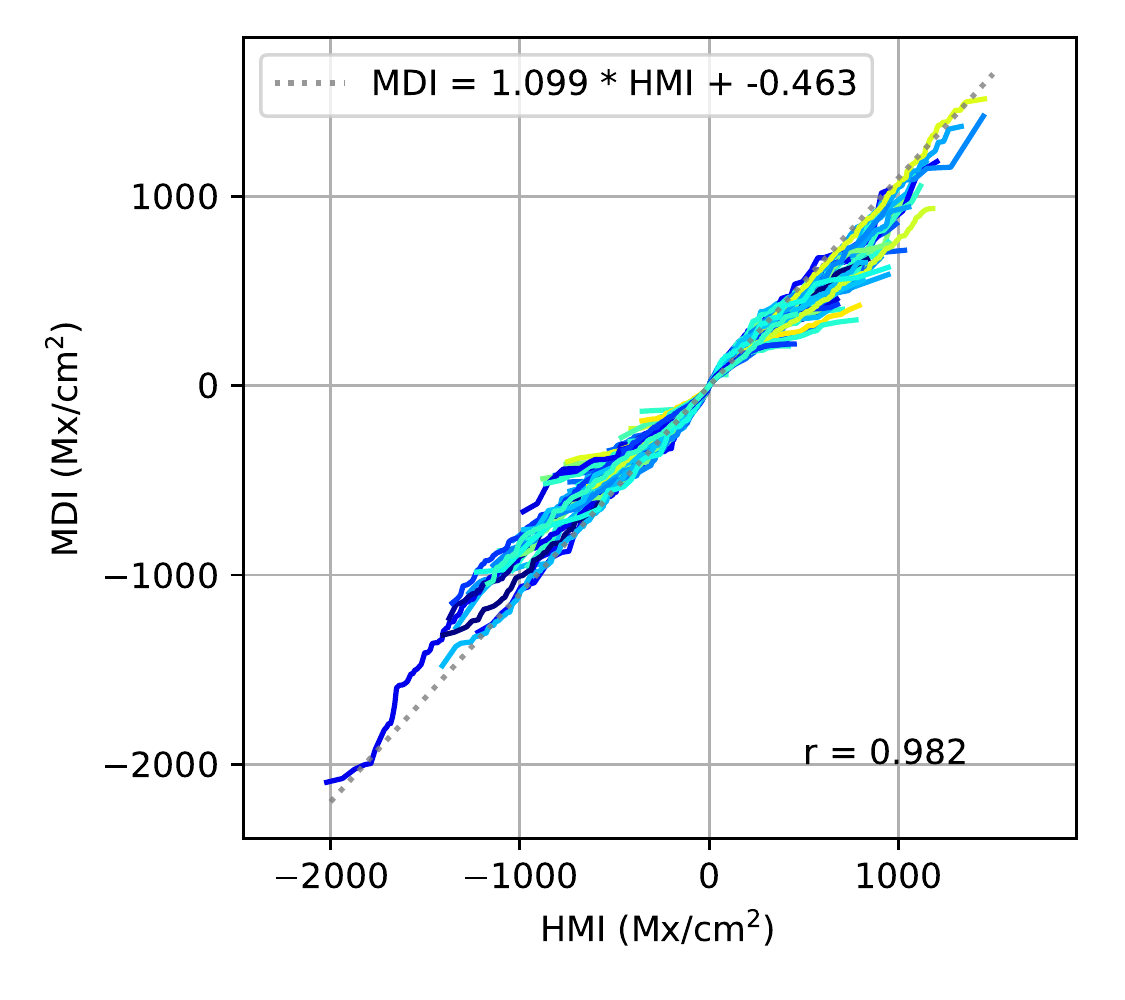}{0.36\textwidth}{}}
    \vspace{-2em}
    \caption{Q-Q (quantile-quantile) plot of 50 matched magnetogram pairs of HARP and TARP from May 1 to October 28 in 2010.
    Active regions with pixels outside of $\pm 70 ^\circ$ from the central meridian are not used.
    For each pair, the co-temporal magnetograms are sampled at a rate of every 8 hours.
    The pixels within the intersection of the bounding boxes of active region pairs are used.
    Lighter colors indicate higher latitudes.
    \label{fig:qq}}
\end{figure}

We first discuss the fusion of the SHARP and SMARP magnetograms. SHARP magnetograms inherit the HMI resolution of about $0.5''$ per pixel, whereas SMARP magnetograms inherit the MDI resolution of about $2''$ per pixel. To compare HMI and MDI magnetograms, \citet{liu2012comparison} reduced HMI spatial resolution to match MDI's by convolving a two-dimensional Gaussian function with an FWHM of 4.7 HMI pixels and truncated at 15 HMI pixels. Then, the HMI pixels enclosed in each MDI pixel are averaged to generate an MDI proxy pixel. Subsequently, a pixel value transformation $\textrm{MDI} = -0.18 + 1.40\times\textrm{HMI}$ is applied.
In this work, we adopted a simpler approach by subsampling SHARP magnetograms 4 times in both dimensions to match the resolution of SMARP magnetograms.
Unlike \citet{liu2012comparison}, we approximated pixel value transformation with an identity map, as pixel value distributions of SHARP and SMARP magnetograms in the overlap period are very similar (Figure \ref{fig:qq}). Our approximation agrees well with \citet{riley2014multi}, who found a multiplicative conversion factor of $0.99\pm 0.13$ between MDI and HMI using histogram equating.
The discrepancy between our multiplicative conversion factor (1.099) and that of \citet{liu2012comparison} (1.40) may be because they considered full-disk magnetograms whereas we focus on active regions. In addition, they considered only 12 pairs in June--August 2010, whereas we considered every possible matching in May--October 2010. Moreover, they performed pixel-to-pixel match of full-disk magnetograms, whereas we use histogram-based methods on active regions because more precise models for aligning coordinates between CEA-projected SHARP and SMARP are not yet available \citep{bobra2021smarps}. Furthermore, they considered pixels within 0.866 solar radius of Sun's center, whereas we considered pixels within $\pm 70 ^\circ$ from the central meridian.

\begin{figure}
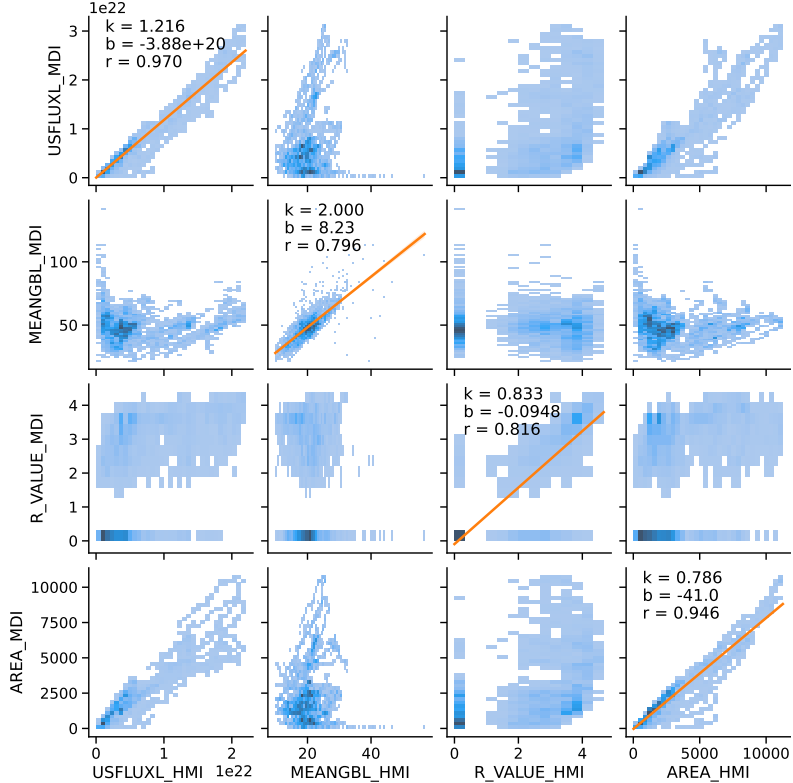

    \centering
    \gridline{
        \fig{eda/target_vs_feature_2.pdf}{0.6\textwidth}{}
    }
    \vspace{-2em}
    \caption{2D histograms of summary parameters \texttt{USFLUXL}, \texttt{MEANGBL}, \texttt{R\_VALUE}, and \texttt{AREA} between SHARP and SMARP. SHARP summary parameters are suffixed with \texttt{\_HMI} and SMARP with \texttt{\_MDI}. The orange lines in the diagonal blocks are the least square fit of SMARP summary parameters on the corresponding SHARP summary parameters, with coefficient $k$, intercept $b$, and Pearson correlation coefficient $r$ displayed in the corner. 
    }
    \label{fig:regplot}
\end{figure}

We next discuss the fusion of SHARP and SMARP summary parameters. Although designed to represent the same physical quantity, summary parameters with identical keywords in SHARP and SMARP are calculated from two  pipelines with different source data, and the differences between them cannot be neglected. \citet{bobra2021smarps} investigated these differences by comparing the marginal and the pairwise joint distribution of co-temporal SMARP and SHARP summary parameters for 51 NOAA active regions over the overlap period of MDI and HMI \citep[][Figure 3]{bobra2021smarps}. Motivated by these findings, we investigated the linear associations between  SHARP and SMARP using a univariate linear regression analysis. Specifically, SMARP parameters were regressed on their SHARP counterparts. As shown in Figure \ref{fig:regplot}, \texttt{USFLUXL} is the most correlated parameter between SHARP and SMARP, with Pearson correlation coefficient $r = 0.970$, whereas \texttt{MEANGBL} is the least correlated parameter, with $r=0.796$.
Note that applying linear transformations to the SHARP summary parameters would have no effect once Z-score standardization was performed. This is because Z-scores are invariant to univariate linear transformation. Therefore, in practice, the linear transformation on SHARP summary parameters is not performed.

\subsection{Sample extraction and labeling}
\label{sec:sample}

\begin{figure}
    \centering
    \includegraphics[width=0.8\textwidth]{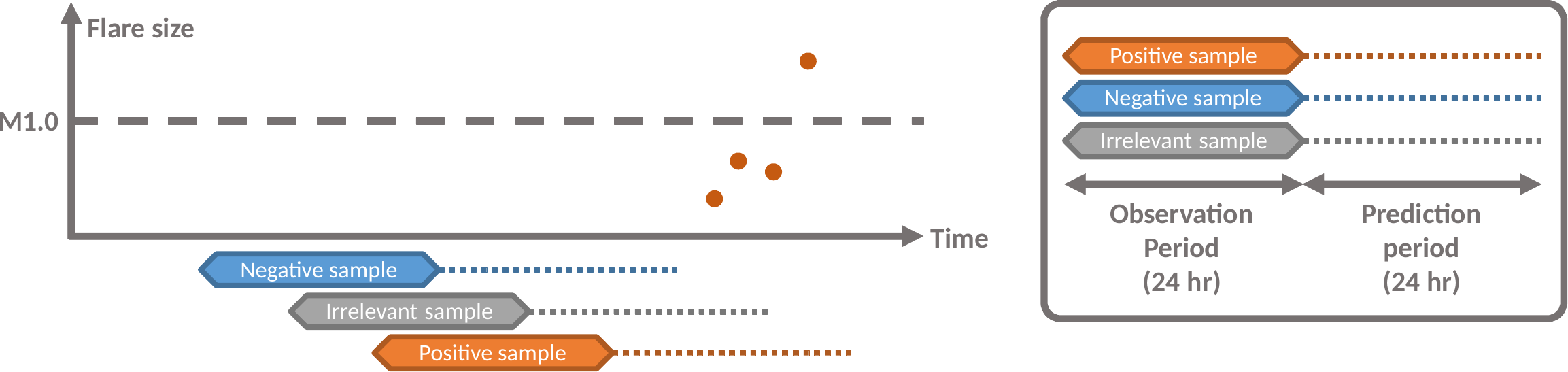}
    \caption{Demonstration of the sample extraction and labeling procedure of an active region. The dark orange dots represent flares that occurred in an active region, with the last flare exceeding the M1.0 threshold. The blue sample is labeled as the negative class because no flare of any class occurs in the observation and the prediction period. The gray sample is irrelevant to the task since all flares in the prediction period are weaker than M1.0. The orange sample is labeled as the positive class because the prediction period contains a flare of size exceeding M1.0.
    }
    \label{fig:sample_def}
\end{figure}

We focus our joint SMARP and SHARP analysis on a particular task of interest, which we refer to as ``strong-vs-quiet" flare prediction: based on a sequence of observations of an evolving active region, the objective is to discriminate active regions that will generate strong flares in the near future, from active regions having no flare activity whatsoever. To construct a dataset for this task, we extract data samples using a sliding window approach similar to \citet{angryk2020multivariate}. Specifically, samples are extracted from a 24-hour time window that slides through an active region sequence with step size of 96 minutes (i.e., one 24-hour subsequence starts 96 minutes after the starting point of its previous subsequence). The 24-hour time window that a sample covers is called the \emph{observation period}, and the 24-hour time window following immediately after the observation period is called the \emph{prediction period}. Then we retain the samples that either: (1) exhibit an M- or X-class flare in the prediction period (assigned to the positive class); or (2) have no flare of any class in both observation and the prediction period (assigned to the negative class). Table \ref{tab:labeling} lists the sample sizes of all the possible flare activity evolution types and the associations between evolution types and class labels. Figure~\ref{fig:sample_def} shows examples of extracted and labeled samples.

We note that two evolution types are excluded in this study. Evolution types denoted by blank spaces in Table~\ref{tab:labeling} indicate a decay in flare activity---a process different from the onset or the continuation of the flare activity. They are unrelated to our task and also less studied in the literature. Including them in the dataset brings an unnecessary source of heterogeneity and makes learning a reliable predictor substantially more difficult. Evolution types denoted by question marks have only weak flares in the prediction period. They are excluded to enhance the contrast between the two classes, to avoid the concerns on the detection of weak flares (many B- and C-class flares are obscured by background radiation), and to avoid the controversy on the granularity of labels (for instance, an M1.0 class flare and a C9.9 class flare relieve a similar amount of energy but are categorized differently). Possible limitations of our sample selection rules are discussed in Section~\ref{sec:discussion}.

\begin{table}
    \centering
    \caption{
    \emph{Left}:
    Samples sizes of all possible flare activity evolution types, with missing/inconsistent data removed. The flare activity in the observation period of a sample can be \texttt{QUIET} (the active region is flare-quiet), \texttt{WEAK} (only flares of size smaller than M1.0 occur), or \texttt{STRONG} (there is at least one large flare of size M1.0 or above). The prediction period can be classified the same way. The entries denote the sample counts in SMARP/SHARP data sets.
    \emph{Right}:
    The associations of flare activity evolution types and the class labels for the ``strong-vs-quiet" flare prediction task.
    Positive samples are denoted as \texttt{+} and negative samples as \texttt{-}. Samples with the evolution type signifying a decaying flare activity (denoted as a blank space) or leading to only small flares (denoted as \texttt{?}) are not relevant to our task.
    }
    \begin{tabular}{cccc}
        \toprule
        & \multicolumn{3}{c}{Prediction} \\
        \cmidrule(lr){2-4}
        & \texttt{QUIET} & \texttt{WEAK} & \texttt{STRONG} \\
        Observation & & & \\
        \midrule
        \texttt{QUIET}  & 130695 / 66349 &  12341 / 10715 &   932 / 296 \\
        \texttt{WEAK}   & 12688 / 11110 &  12033 / 14891 &  1915 / 1366 \\
        \texttt{STRONG} & 1071 / 282 &   1723 / 1371 &  1754 / 1187 \\
        \bottomrule
    \end{tabular}
    \begin{tabular}{cccc}
        \toprule
        & \multicolumn{3}{c}{Prediction} \\
        \cmidrule(lr){2-4}
        & \texttt{QUIET} & \texttt{WEAK} & \texttt{STRONG} \\
        Observation & & & \\
        \midrule
        \texttt{QUIET}  & \texttt{-} & \texttt{?} & \texttt{+}\\
        \texttt{WEAK}   & \texttt{ } & \texttt{?} & \texttt{+}\\ 
        \texttt{STRONG} & \texttt{ } & \texttt{ } & \texttt{+}\\
        \bottomrule
    \end{tabular}
    \label{tab:labeling}
\end{table}


After extracting and labeling active region samples, we discarded the samples having inconsistent or missing data. We consider a point-in-time record in a sample sequence as a ``bad record" if (1) the magnetogram contains Not-a-Number (NaN) pixels, (2) the magnetogram has either height or width deviating more than 2 pixels from the median dimension in the sample sequence, or (3) any of the summary parameters is NaN. A sample is discarded if it contains more than 2 bad records or the last record is a bad record. The validity of the last record is enforced because the CNN uses only the last record in the sample sequence. 

The numbers of SMARP and SHARP samples output by the above pipeline are shown in Table \ref{tab:tally}.
The count of negative samples is observed to dominate in both SMARP and SHARP. To address the issue of significant class imbalance, we randomly undersample the negative samples to equalize the positive and negative classes, which will be described in more detail in Section \ref{sec:rus}.

\begin{table}
    \centering
    \caption{Sample sequences extracted from SMARP and SHARP}
    \begin{tabular}{cccc}
        \toprule
         & Positive (M1.0+) & Negative (Quiet) & Event Rate\\
        \midrule
        SMARP & 4601 & 130695 & 0.0340 \\
        SHARP & 2849 & 66349 & 0.0412 \\
        \bottomrule
    \end{tabular}
    \label{tab:tally}
\end{table}

\subsection{Train/validation/test split}


A common practice in machine learning is to divide the data samples into three disjoint subsets, as known as splits: a training set on which the model is fitted, a validation set on which hyperparameters are selected, and a test set on which the model is evaluated for generalization performance. 
The ability of a trained machine learning algorithm to generalize to unseen samples hinges on the distributional similarity among splits. Therefore, it is important that splits be sufficiently similar in distribution. 

Due to the temporal coherence of an active region in its lifetime, a random split of data samples will have samples coming from one active region categorized into different splits. Such correlation constitutes an undesirable information leakage among splits. For instance, information leaking from the training set into the test set will likely result in an overly optimistic estimate of the generalization performance. 
Much of the flare prediction literature deals with this issue by taking a chronological split, e.g., a year-based split \citep[e.g.][]{bobra2015solar, chen2019identifying}. Unfortunately, it is observed that the splits may not share the same distribution due to solar cycle dependency \citep{wang2020predicting}. Some other works take an active-region-based split, where data samples from the same active region must belong to the same split \citep[e.g.][]{guerra2015ensemble, campi2019feature, zheng2019solar, li2020predicting}. Compared to splitting by years, this approach has the advantage that active regions in each split are randomly dispersed in different phases of a solar cycle, removing the bias introduced by artificially specifying splits. This distributional consistency between splits comes at the price of an additional source of information leakage due to sympathetic flaring in co-temporal active regions.

\subsection{Random undersampling}
\label{sec:rus}

As shown in Table \ref{tab:tally}, both SMARP and SHARP exhibit prominent class imbalance, with positive (minority class) samples significantly outnumbered by negative (majority class) samples.
In flare forecasting, class imbalance has been recognized as a major challenge, both in forecast verification \citep{woodcock1976evaluation,jolliffe2012forecast} and in data-driven methodology \citep{bobra2015solar,ahmadzadeh2021train}.
A data-driven forecasting method needs to be calibrated to handle class imbalance properly, in order to effectively detect the events of interest, as opposed to being overwhelmed by the sheer volume of the negative samples in the training set.

Methods to tackle class imbalance can be categorized into three types: data-level methods, algorithm-level methods, and a hybrid of the two \citep{krawczyk2016learning,johnson2019survey}. 
Data-level methods rebalance the class distribution by oversampling the minority class and/or undersampling the majority class---both have been used in flare forecasting \citep[e.g.][]{ribeiro2021machine,yu2010short}. Classifiers trained on rebalanced datasets, without being biased towards the majority class, are more likely to effectively detect the event of interest. Such classifiers are also generally more robust to variations in class imbalance than classifiers trained on the original imbalanced data \citep{xue2014does}.
Algorithm-level methods modify the learners to alleviate their bias towards the majority groups. The most popular algorithm-level approach---also widely used in flare forecasting \citep[e.g.][]{bobra2015solar, nishizuka2018deep, liu2019predicting}---is cost-sensitive learning, which assigns a higher penalty to the misclassification of samples from the minority class to boost their importance \citep{krawczyk2016learning}. The penalty weights for different classes are usually set to be inversely proportional to their samples sizes \citep{nishizuka2018deep,liu2019predicting,ahmadzadeh2021train}.
Other algorithm-level approaches include imbalanced learning algorithms, one-class learning, and ensemble methods \citep{ali2013classification}, but they are rarely used in flare forecasting. Recent work by \citet{ahmadzadeh2021train} provided a thorough investigation of class imbalance in flare forecasting by presenting an empirical evaluation of multiple approaches.

We handle the class imbalance problem using random undersampling: we randomly remove samples from the majority class until the number of positive and negative samples are equalized. By training on rebalanced training and validation sets, we obtain a predictor that is more robust to shifts in climatological flare rates and learns more resilient pre-flare features. Following \citet{zheng2019solar} and \citet{deng2021fine}, we also perform random undersampling on test sets to preserve the class proportion consistency among splits.
By testing on rebalanced test sets, we evaluate the generalization performance of the predictor under the \emph{same} climatological rate as it is trained with. We note there are also studies that do not rebalance the test set in order to evaluate the performance under a realistic event rate \citep{cinto2020framework,ahmadzadeh2021train}. However, the bias caused by the class-balance change between the test and the training set is often neglected. We discuss such bias in addition to possible corrective methods in Section~\ref{sec:discussion}. Applying such corrections will be left to future work that directly addresses operational applications.

\subsection{Image resizing}

The CNN requires all input images to be of the same size, but the active region cutouts are of different sizes and aspect ratios. Resizing (via interpolation), zero padding, and cropping are among mostly used methods to convert different-sized images into a uniform size. \citet{jonas2018flare} cropped and padded input images to a square aspect ratio and then downsampled them to $256\times 256$ pixels. This has the advantage of preserving the aspect ratio. However, since many active regions are east-west elongated, cropping may exclude part of active regions and padding may introduce artificial signals. In this work, we resize all active region magnetograms to $128\times 128$ pixels using bilinear interpolation, similar to \citet{huang2018deep} and \citet{li2020predicting}.


\subsection{Standardization}

Magnetogram pixel values and summary parameters are different physical quantities expressed in different units and ranges.
Unlike physical modeling, many machine learning algorithms are invariant to scaling the input; they only care about the relative feature amplitudes. 
Moreover, drastically different ranges of features may hurt the convergence and stability of many algorithms. Therefore, the data of different scales are typically transformed into the same range via a process called standardization. In particular, Z-score standardization transforms the input data by removing the mean and then dividing by the standard deviation.


In this work, we apply the Z-score standardization to the image data using the mean and standard deviation of the magnetogram pixels in SHARP.
This is because the pixel values between SMARP and SHARP are similar. We apply the Z-score standardization to SMARP and SHARP summary parameters separately. That is, the mean and standard deviation are calculated for SHARP and SMARP separately, and data in one dataset is standardized using the mean and the standard deviation in that dataset. The transformation is ``global" \citep{ahmadzadeh2021train} in that it is calculated regardless of the splits. Empirical evaluation in \citet{ahmadzadeh2021train} showed a global standardization to be better than the local standardization, i.e., the mean and standard deviation are calculated only for the training split. We note that, with this standardization, the linear transformation converting SHARP summary parameters to SMARP proxy data is no longer needed; any coefficients and bias will have no effect after standardization.


\section{Methodology}
\label{sec:methodology}

In this section, we introduce the two deep learning models, LSTM and CNN, that we use for flare prediction. Then we describe the stacking ensemble approach that combines the two models. Subsequently, we describe the forecast verification methods (skill scores and graphical tools) we used. Then we discuss how we use the paired $t$-test to compare empirical performance between algorithms and settings with statistical confidence. Lastly, we introduce the visual attribution methods used to interpret the decisions generated by the CNN.

\subsection{Deep learning models}

We use two deep neural network models, LSTM and CNN, to predict strong flares from active region observation. LSTMs use 24-hour-long time series of summary parameters before the prediction period begins, whereas CNNs use the static point-in-time magnetogram right before the prediction period begins. Both networks output the probability that an input sample belongs to the positive class, i.e., the probability that the active region will produce a strong flare in the next 24 hours, rather than continue to be flare-quiescent.

\begin{figure}
    \gridline{\fig{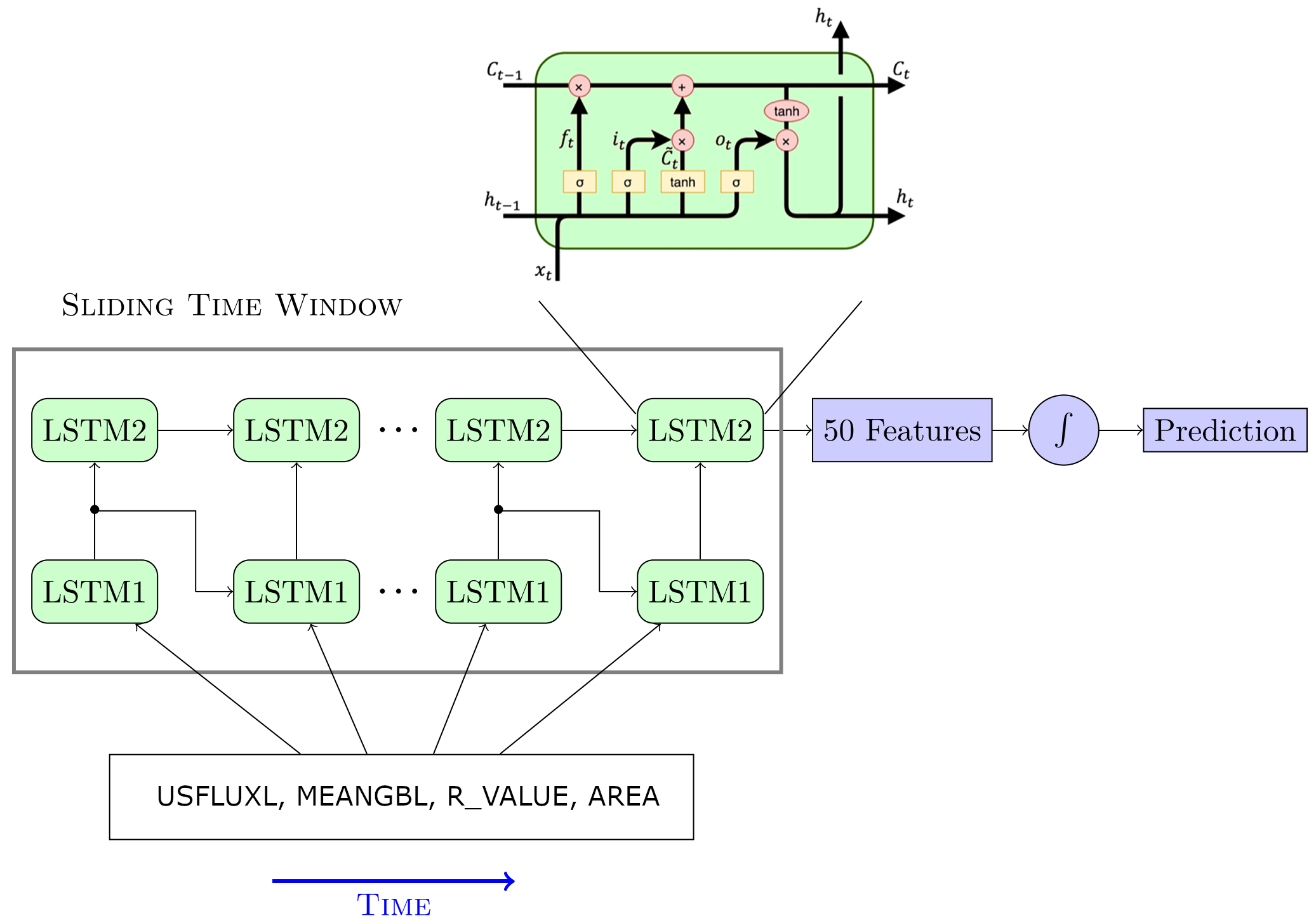}{0.6\textwidth}{(a) LSTM architecture}
              }
    \gridline{\fig{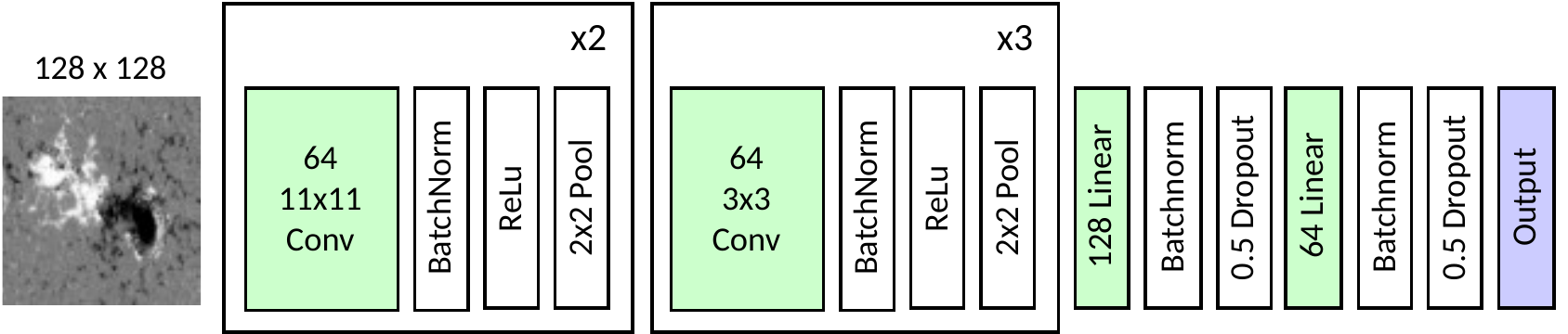}{0.7\textwidth}{(b) CNN architecture}
              }
    \caption{Neural network architectures. (a) shows the LSTM architecture. (b) shows the CNN architecture.}
    \label{fig:arch}
\end{figure}

The long short-term memory (LSTM) network was introduced by \citet{hochreiter1997long} as a type of recurrent neural networks that learns from sequential data for classifying text and speech. A common LSTM unit is composed of a cell, an input gate, an output gate, and a forget gate. In solar flare prediction, LSTMs have been applied to prediction using SHARP parameter series \citep{chen2019identifying, liu2019predicting}. The architecture of the LSTM used in this paper is adapted from \citet{chen2019identifying}, shown in Figure \ref{fig:arch}(a). Two LSTM layers, each with 64 hidden states, are stacked. The last output of the second LSTM layer, a 64-dimensional vector, is sent to a linear layer with 2 outputs. The softmax is applied to this 2-dimensional output to get the predicted probabilities of the positive and the negative class.

The convolutional neural network (CNN) is a neural network architecture that learns from images. CNNs have been applied to solar flare forecasting by \citet{huang2018deep} and \citet{li2020predicting}. We use the architecture proposed by \citet{li2020predicting}, illustrated in Figure \ref{fig:arch}(b), which was itself inspired by the VGG network \citep{simonyan2014very} and the Alexnet network \citep{krizhevsky2012imagenet}. The first two convolutional layers have kernels of size $11 \times 11$, designed to learn low-level and concrete features. The three following convolutional layers have kernels of size $3\times 3$, designed to learn more high-level, abstract concepts. Batch normalization is used after all convolutional and linear layers to speed convergence. ReLU nonlinearity is applied to only convolutional layers. The batch normalization outputs of the two linear layers are randomly dropped out with a probability of 0.5 in training to reduce overfitting. The 2-dimensional output is passed to softmax to generate a probability assignment between the positive and the negative class. More details of this architecture can be found in \citet{li2020predicting}.

The procedures used to train the LSTM and the CNN are similar. For both models, the Adam optimizer \citep{kingma2014adam} is used to minimize the cross-entropy loss with learning rate $10^{-3}$ and batch size 64. Both models are evaluated on the validation set after each epoch of training. To prevent overfitting, the training is early-stopped if no improvement on the validation True Skill Score (or TSS, explained later in Section \ref{sec:evaluation}) is observed for a certain number of epochs, called the \emph{patience},  before early stopping. The LSTM is trained for at most 20 epochs with a patience of 5 epochs, whereas the CNN is trained at most 20 epochs with a patience of 10 epochs. After training, the LSTM or the CNN with the best validation TSS among the checkpoints of all epochs is selected and evaluated on the test set to estimate its generalization performance.

\subsection{Stacking ensemble}

In ensemble learning, the most common approach to combining individual models---called \emph{base learners}---is averaging their outputs, possibly with non-equal weights, to produce a final output.
Another approach is stacking. Different from output averaging, stacking uses cross-validation training to learn the best combination---called the \emph{meta-learner}---of the outputs of the base learners. The meta-learner is often chosen to be global and smooth \citep{wolpert1992stacked}, such as linear models \citep{breiman1996stacked,leblanc1996combining,ting1999issues} and decision trees \citep{todorovski2003combining,dvzeroski2004combining}. Training a stacking ensemble consists of two stages. First, the base learners are fitted on the training set. Then, the predicted probabilities by all base learners on the validation set, as well as their labels, are collected into the so-called \emph{level-one} data, on which the meta-learner is fitted. Cross-validation is frequently used in place of a simple train-validation split to significantly increase the sample size of the level-one dataset. Either way, it is important that the level-one data are out-of-sample for base learners, otherwise the meta-learner will inevitably prefer models that overfit the training data over ones that make more realistic decisions \citep{witten2016data}.

An early application of stacking for solar flare prediction problems was presented by  \citet{dvzeroski2004combining}. These authors proposed a decision-tree-based stacking method, demonstrating it on the {\it UCI Repository of machine learning databases} \citep{Dua:2019}, including a dataset with 1389 flare instances, each characterized by 10 categorical attributes.
In the space weather community, \citet{guerra2015ensemble} proposed stacking for flare prediction. They linearly stacked four full-disk probabilistic forecasting methods, with the weights maximizing HSS under the constraint that they are non-negative and sum to 1.
They found that the stacking ensemble performed similarly to an equally weighted model.
\citet{guerra2020ensemble}  continued in this direction adopting stacking for more forecasting methods and a larger data sample. They also considered an unconstrained linear combination with a climatological frequency term. They found most ensembles perform better than a bagging model that essentially averages the members' predictions. However, these authors overlooked the nonconvexity of the objective in training the meta-learner.
Furthermore, their conclusions about the superiority of the stacking ensemble over the equally weighted model were limited to the evaluation of in-sample error, which is unlikely to generalize. We will discuss these issues as we introduce our proposed stacking ensemble.



We formulate the stacking ensemble as a linear combination of LSTM and CNN. Given a sample $x_i$, the stacking ensemble outputs the probability that the sample belongs to the positive class
\begin{align}
    r_i = \alpha p_i + (1-\alpha) q_i,\quad 0\leq\alpha\leq 1, \label{eq:meta}
\end{align}
where $p_i$ and $q_i$ are the predicted probability by LSTM and CNN, respectively, and $\alpha$ is the meta-learner parameter. We note that, in order to be most effective, stacking methods require base learners that are diverse and complement each other. Examples include base learners trained on different types of data, each providing an alternative view of the same phenomenon. The magnetograms and summary parameters in SHARP/SMARP provide such diverse multiviews. These multiviews are processed by CNN and LSTM, respectively, to generate two predicted probabilities, which are then fused into a single prediction by the aforementioned stacking procedure.

The meta-learning of the stacking ensemble essentially means finding the optimal combination weight $\alpha$. To that end, we minimize a loss function that penalizes the difference between the prediction $r_i$ and the binary label $y_i\in\{0, 1\}$ for samples in the validation set. A natural choice is to maximize the accuracy or skill scores, or equivalently, to minimize the loss which is the negation of these metrics.
The downside of these loss functions is that they may not be convex or differentiable. This is not problematic when there are only a few base learners, as is the case with this work and \citet{guerra2015ensemble}; a grid search can be applied to find the weights that minimize the loss. However, as the number of base learners increases, the grid search quickly becomes infeasible, and iterative algorithms have to be used---many of them require convexity and differentiability of the loss function for guaranteed convergence. In machine learning, convexity and smoothness ensure the uniqueness of the minimizer and guarantee faster convergence rates for iterative algorithms \citep{nocedal2006numerical,bottou2018optimization}. \citet{guerra2020ensemble} found that for some optimization metrics, the resulting weights were sensitive to the initialization of the solver. This is likely the consequence of the nonconvexity of the loss function. Their proposed solution was to run the solver with multiple initializations and take the average.

To circumvent convergence problems, machine learning researchers often use loss functions that are convex and differentiable. One example is the negative log-likelihood function for the logistic regression model, whose minimum corresponds to the maximum likelihood estimator (MLE). Within the meta-learning framework specified by Equation~(\ref{eq:meta}), 
the negative log-likelihood loss function is
\begin{align}
    L(\alpha) &= -\log \prod_{i=1}^n r_i^{y_i} (1-r_i)^{1-y_i} \\
    &= \sum_{i=1}^n \underbrace{\left( -y_i \log r_i - (1 - y_i) \log (1-r_i) \right)}_{L_i}\,.
\end{align}
The negative log-likelihood objective can also be interpreted as the binary cross-entropy loss, a divergence measure between the distributions of ground truth labels and predicted probabilities. This loss function can be decomposed into the summation of instance-wise loss $L_i$, having gradient and the Hessian
\begin{align}
    L_i'(\alpha) &= \left(-\frac{y_i}{r_i} + \frac{1-y_i}{1-r_i}\right)(p_i - q_i)\,, \\
    L_i''(\alpha) &= \left(\frac{y_i}{r_i^2} + \frac{1-y_i}{(1-r_i)^2}\right) (p_i - q_i)^2 \geq 0.
\end{align}
Minimizing $L$ on $\alpha\in[0, 1]$ is a convex optimization problem and we use grid search to find the minimizer. In general cases with more than two base learners, iterative algorithms like projected gradient descent or Newton's method will be more efficient. We point out that convex loss functions are widely adopted in the literature of stacking, such as least square estimate \citep{breiman1996stacked}, regularized linear regression \citep{leblanc1996combining}, multi-response linear regression \citep{ting1999issues}, and hinge loss \citep{csen2013linear}. 

\subsection{Evaluation tools}
\label{sec:evaluation}


The prediction probabilities output by CNN and LSTM can be turned into binary decisions by thresholding and the algorithm performance can be represented as a contingency table (or confusion matrix), as shown in Table \ref{tab:contingency}. The contingency table contains the most complete information for categorical prediction. However, a single numerical metric is often needed to summarize the table for model selection.
Accuracy and skill scores are examples of such contingency table based metrics that are adopted in space weather forecasting.

\begin{table}
    \centering
    \renewcommand\arraystretch{1.2}
    \settowidth\rotheadsize{\theadfont True}
    \begin{tabular}{@{} cc|c|c|c}
        \multicolumn{2}{c}{} & \multicolumn{2}{c}{Predicted} & \\
        & & Negative & Positive & Total\\
        \cline{2-4}
        \multirow{2}{*}[0ex]{\rothead {True}}
        & Negative & TN & FP & $\Ntrue$ \\
        \cline{2-4}
        & Positive & FN & TP & $\Ptrue$ \\
        \cline{2-4}
        & \multicolumn{1}{c}{Total} & \multicolumn{1}{c}{$\Npred$} & \multicolumn{1}{c}{$\Ppred$} & \multicolumn{1}{c}{$\Total$}
    \end{tabular}
    \hspace{70px} 
    \vspace{2ex}
    
    
    \caption{A contingency table consisting of TP (true positive), FP (false positive), FN (true negative), and TN (true negative).}
    \label{tab:contingency}
\end{table}

We start our discussion on metrics with accuracy (ACC), also known as rate correct, the simplest metric that is widely used in all sorts of domains. In terms of the contingency table, accuracy is defined as 
\begin{align}
    A = \frac{\TN + \TP}{\Ntrue + \Ptrue}.
\end{align}
For a highly imbalanced classification problem like solar flare prediction, accuracy is generally not considered a useful metric, since a no-skill classifier that assigns the majority label to all samples will be correct most of the time. Therefore, a plethora of skill scores are devised to overcome this issue.




A skill score provides a normalized measure  of  the  improvement  against  a  specific  reference method. In its most general form, a skill score can be expressed as
\begin{align}
    \textrm{Skill} = \frac{A_{\textrm{forecast}} - A_{\textrm{reference}}}{A_{\textrm{perfect}} - A_{\textrm{reference}}},
\end{align}
where $A_{\textrm{forecast}}$, $A_{\textrm{reference}}$, and $A_{\textrm{perfect}}$ are the accuracy of the forecast to be evaluated, the reference forecast, and the perfect forecast, respectively.
A higher skill score indicates better performance, with the maximum value 1 corresponding to the perfect performance, 0 corresponding to no improvement over the reference, and negative values corresponding to performance worse than the reference. Below, we review some of the mostly used skills scores in flare forecasting. For a more complete discussion, we refer readers to \citet{woodcock1976evaluation} and \citet{wilks2011statistical}.


The Heidke Skill Score (HSS), also known as Cohen's kappa coefficient due to \citet{cohen1960coefficient}, uses a random forecast independent from the flare occurrences as a reference. The expected number of correct forecasts made by the random predictor, denoted by $\textrm{E}$, can be calculated using the law of total expectation as
\begin{align}
    \textrm{E} = \frac{\Ptrue}{\Total} \times \Ppred + \frac{\Ntrue}{\Total} \times \Npred.
\end{align}
The accuracy of the random predictor can then be expressed as 
\begin{align}
    A_{\textrm{reference}} = \frac{\textrm{E}}{\Total}.
\end{align}
Defined using this reference accuracy, $\HSS$ has the form
\begin{align}
    \HSS = \frac{\TP + \TN - \textrm{E}}{\Total - \textrm{E}} = \frac{2[(\TP\times\TN) - (\FN\times\FP)]}{\Ptrue\Npred + \Ppred\Ntrue}.
\end{align}
HSS quantifies the forecast improvements over a random prediction. Since the random reference forecast is dependent on the event rate (climatology) $\Ptrue / (\Total)$, HSS has to be used with discretion in comparing methods when the event rate varies.


The True Skill Score (TSS), also known as Hanssen \& Kuiper's Skill Score (H\&KSS) or Peirce Skill Score. It is the difference between the probability of detection (POD) and the false alarm rate (FAR):
\begin{align}
    \TSS = \underbrace{\frac{\TP}{\Ptrue}}_{\text{POD}} - \underbrace{\frac{\FP}{\Ntrue}}_{\text{FAR}}.
\end{align}
TSS falls into the general skill score definition with a reference accuracy \citep{barnes2016comparison}
\begin{align}
    A_{\textrm{reference}} = \frac{\FN(\TN - \FP)^2 + \FP(\TP + \FN)^2}{(\Ntrue + \Ptrue)[\FN(\TN - \FP) + \FP(\TP + \FN)]}\,,
\end{align}
constructed such that both the random and unskilled predictors score 0. A nice property of TSS is its invariance to the class imbalance ratio, and hence is suggested by \citet{bloomfield2012toward} to be the standard measure for comparing flare forecasts.


We note that, on a balanced dataset for which the event rate is 0.5, it can be shown that $\TSS = \HSS = 1 - 2(1 - \ACC)$. The trend and the paired $t$-test results for $\TSS$ apply to $\ACC$ and $\HSS$ due to perfect correlation. Therefore, we mainly focus on the discussion on $\TSS$, list $\ACC$ as a complement metric, and omit $\HSS$ as it is equal to $\TSS$ in our setting.
For probabilistic forecasts, the aforementioned metrics (ACC, HSS, and TSS) depend upon the threshold applied to the predicted probability. A common practice is to apply a threshold of 0.5, which is considered to be ``random" by many researchers. In contrast, the following two metrics, BSS and AUC, are irrelevant to the threshold, and they need information (i.e., predicted probabilities) beyond the mere contingency table.

The Brier Skill Score (BSS) is a skill score evaluating the quality of a probability forecast. It is of a nature different from those of HSS and TSS, in that it directly uses probabilistic predictions without thresholding them. The BSS also admits the general skill score formulation, with the accuracy replaced by the Brier Score (BS), defined as the mean squared error between the probability predictions $f_i$'s and binary outcomes $o_i$'s:
\begin{align}
    \textrm{BS} = \frac{1}{n}\sum_{i=1}^n (f_i - o_i)^2\,.
\end{align}
With a reference forecast that consistently predicts the average event frequency $\bar{o}$ (also known as climatology), the $\BSS$ is given by
\begin{align}
    \BSS = 
    1 - \frac{\BS_\textrm{forecast}}{\BS_\textrm{reference}}\,.
\end{align}
It is sometimes of interest to decompose $\BS$ into three components of reliability, resolution, and uncertainty \citep{murphy1973new, mccloskey2018flare}. BSS is frequently accompanied by the reliability diagram discussed below, providing more complete information about the performance of probabilistic predictions.

For completeness we briefly discuss the Area Under Curve (AUC), defined as the area under the receiver operating characteristic (ROC) curve. The ROC curve depicts how the probability of detection changes with the false alarm rate by varying the classification threshold. A higher AUC implies a higher average detection probability over all false alarm rates. Unlike dichotomous metrics like TSS and HSS, AUC summarizes detection performance over all possible false positive rates and, in particular, does not depend on the threshold selected to convert probabilistic forecasts into binary decisions. 
Consequently, the AUC is not as useful in flare prediction, especially when stringent false positive control is exercised \citep{steward2017automatic}. 

The above skill scores provide one way to directly compare flare prediction models. In addition to such metrics, flare forecasts are often evaluated using graphical tools 
for diagnostics and comparison. Apart from ROC curves discussed above, reliability diagrams (RD) and skill score profiles (SSP) are also commonly used in forecast verification. All three of them are applicable to forecasts that predict probabilities or continuous scores (e.g., logits) that can be converted to probabilities. We briefly discuss RD and SSP below.

\begin{figure}
    \centering
    \includegraphics[width=0.43\textwidth]{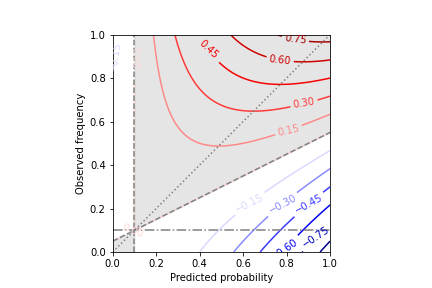}
    \caption{An illustration of the relation between the reliability diagram and BSS.}
    \label{fig:bss}
\end{figure}

The reliability diagram, also known as the calibration curve, measures how well a probabilistic forecast agrees with the observation. The predicted probabilities are binned into groups and the observed event rate within each group is plotted. If the predicted probability agrees well with the observed rate, the points will be close to the diagonal of the plot (the line of perfect reliability). Such a forecast is called reliable. Any forecast that produces predictions independent of flare activity has all its points close to the horizontal line at the event rate.
BSS provides a metric that accounts for both reliability and resolution. Figure \ref{fig:bss} shows an example of the plane on which the reliability diagram is drawn. The climatology rate is set to be 0.1. The overall BSS can be seen as a histogram weighted average of the contributions of the points on the reliability diagram. The contours are equal contribution lines. The points in the shaded area contribute positively to BSS. The dashed line with slope 1/2 is called the ``no skill" line, the points on which have zero contribution to the overall BSS.

A skill score profile plot shows how skill scores change as a function of the probability threshold. A method with high and flat profile is usually desired, as such a method achieves high skill scores and the performance is robust to the changes of the threshold.


\subsection{Statistical performance comparisons}
We use a one-sided paired $t$-test to assess the comparative performance of a pair of prediction algorithms, called algorithm 1 and 2, that are tested on the same test data.
Specifically, two competing hypotheses are formulated: the null hypothesis ($H_0$) that algorithms 1 and 2 have identical performance and the alternative hypothesis ($H_1$) that algorithm 2 is better than algorithm 1.
Suppose we have $n$ pairs of empirical performance $\{(x_i,y_i)\}_{i=1}^n$ achieved by algorithm 1 and 2 on $n$ test samples.
The paired $t$-statistic is $t=\sqrt{n}(\overline{y}-\overline{x})/\sigma$ where $\overline{y}$ and $\overline{x}$ are the sample means of $\{x_i\}_{i=1}^n$ and $\{y_i\}_{i=1}^n$, respectively, and $\sigma^2$ is the sample variance of the differences $\{y_i-x_i\}_{i=1}^n$.
Under $H_0$, $t$ follows a Student-$t$ distribution with $n-1$ degrees of freedom  \citep{bickel2015mathematical}. The $p$-value associated with the test statistic $t$ is defined as the area under the Student-$t$ density to the right of the value $t$. Small $p$-values provide strong evidence in favor of $H_1$, i.e., that algorithm 2 is better than algorithm 1. 
In this paper, we use the paired $t$-test to examine the following hypotheses:
\begin{itemize}
    \item Training a predictor (LSTM or CNN) on data from two solar cycles (SMARP and SHARP) improves upon training a predictor on data from a single solar cycle (only SMARP or only SHARP) (Section~\ref{sec:results_data}).
    \item LSTM achieves better performance than CNN (Section~\ref{sec:results_model}).
    \item The LSTM-CNN stacking ensemble achieves better performance than the better model between LSTM and CNN (Section~\ref{sec:results_stacking}).
\end{itemize}

\subsection{Interpretation of CNNs} 

Deep learning methods 
are widely applied to many domains such as computer vision, natural language processing, speech processing, robotics, and games \citep[see, e.g.][]{he2016deep, silver2016mastering, devlin2018bert}. As of today, deep learning algorithms largely remain black box methods, raising concerns of lack of interpretability, transparency, accountability, and reliability. Interpretability is of particular importance 
when deep learning is used in scientific discovery. Over the years, many tools for interpreting the functioning of deep neural networks have been proposed, revealing aspects of their underlying decision process.





One way to interpret a black-box model, often referred to as ``attribution", is to see how different parts of the input contribute to the model's output. An attribution method generates a vector of the same size as the input, with each element indicating how much the corresponding element in the input contributes to the model decision for that input. In the context of CNNs, the attribution vector is a heatmap of the same size as the input image.

A multitude of attribution methods have been proposed for CNNs in the task of image classification. One type of approach is perturbation-based methods, among which occlusion \citep{zeiler2014visualizing} is well known. Occlusion masks the input image with a gray patch at different locations and sees how much the prediction score of the ground truth class drops. The prediction score drop varies with location, forming a heatmap, with large values indicating the positions of the features important to the CNN's correct prediction. One drawback of the occlusion method is that it is computationally expensive. Another drawback is that the attribution depends on the size and shape of the patch, which need to be tuned to obtain sensible results. Therefore, this type of approach is not used in our work.

Another type of approach uses gradient-based methods, the basic idea being that the gradient of the predicted score of a certain class with respect to the input reveals the contribution of each dimension of the input. Saliency map \citep{simonyan2013deep}, one of the earliest gradient-based methods, is simply the absolute value of the gradients. The intuition is that the magnitude of the derivative indicates which pixels need to be changed the least to affect the class score the most \citep{simonyan2013deep}.
Deconvolution Network \citep{zeiler2014visualizing} and Guided Backpropagation \citep{springenberg2015striving} attribution methods modify the backpropagation rule. 
Integrated Gradients \citep{sundararajan2017axiomatic} integrate the gradients along the path from a reference image to the target image. Formally, the integrated gradient along the $i$-th dimension for an input $x$ and a baseline $x'$ is
\begin{align}
    L^c_i(x; x') = (x_i - x_i') \times \int_{\alpha=0}^1 \frac{\partial F_c(x' + \alpha \times (x - x'))}{\partial x_i}\,\mathrm{d}\alpha,
\end{align}
where $F_c(x)$ is the model output for class $c$ with input $x$. One desirable property of Integrated Gradients, known as completeness, is that the pixels in the attribution map add up to the difference of prediction scores of the target and the reference image, i.e., $F(x) - F(x')$. DeepLIFT \citep{shrikumar2017learning} and its gradient-based interpretation \citep{ancona2018towards} can be seen as the gradient with modified partial derivatives of non-linear activations with respect to their inputs. Grad-CAM (Gradient-weighted Class Activation Mapping) \citep{selvaraju2017grad} accredits decision-relevant signatures by generating a saliency map, highlighting pixels in the input image that increase the confidence of the network's decision for a particular class.
More formally, the Grad-CAM heatmap $L^c$ for class $c$ with respect to a particular convolutional layer is given by the positive part of the weighted sums of the layer's activation maps $A_k$, i.e.,
\begin{align}
    L^c &= \textrm{ReLU}\left(\sum_k \alpha_k^c A^k\right)\,,
\end{align}
with weights $\alpha^c_k$ given by the spatial average of partial derivatives of the class-specific score $y^c$ with respect to the class activation map as
\begin{align}
    \alpha_k^c &= \frac{1}{Z} \sum_i \sum_j \frac{\partial y^c}{\partial A^k_{ij}}\,,
\end{align}
where $Z$ is a normalization constant. Intuitively, a class activation map is weighted more if the pixels therein make the CNN more confident in its decision that the input belongs to class $c$.

In solar flare prediction, \citet{bhattacharjee2020supervised} applied the occlusion method and found that CNNs pay attention to polarity inversion regions. \citet{yi2021visual} applied Grad-CAM to CNNs and found that polarity inversion lines in full-disk MDI and HMI magnetograms are highlighted as an important feature for flare prediction. In this paper, using a variety of attribution methods, we observe similar trends for the CNN trained on SHARP and SMARP data.




\section{Results}
\label{sec:results}
We aim to answer the following four scientific questions:
(1) Can additional data from another solar cycle benefit the performance of deep learning methods for solar flare prediction?
(2) Do features implicitly learned by CNN work better than handcrafted physical parameters used by LSTM?
(3) Can we combine the two deep learning methods to obtain a better prediction?
(4) What preflare signatures can the CNN detect from the magnetogram of an active region?

To summarize, we report the following findings:
(1) Additional training data from HMI collected in Solar Cycle 24 improve the predictive performance of both LSTM and CNN when tested on Solar Cycle 23.
(2) LSTM (using summary parameters) generally outperforms CNN (directly using magnetograms) in flare prediction.
(3) Stacking CNN and LSTM generally leads to better prediction performance.
(4) Visual attribution methods help us interpret the decision of CNN by identifying preflare features.
This section presents the empirical results that lead to these findings.



\subsection{Data from another solar cycle improves prediction}
\label{sec:results_data}


A major goal of this paper is to examine the utility of using SMARP and SHARP together. We set an experimental group and a control group and contrast their 24-hour ``strong-vs-quiet" flare prediction performance. The control group consists of models that train, validate, and test exclusively on SHARP data. We refer to this type of dataset as \sharponly{}. Compared to the control group, models in the experimental group have the training set enriched by SMARP data, while the validation and the test set are kept the same. We call this type of dataset \fusedsharp{}. The only difference between \sharponly{} and \fusedsharp{} is that models using \fusedsharp{} have access to data from a previous solar cycle in the training phase. Symmetrically, we design \smarponly{} and \fusedsmarp{} to examine the utility that SHARP brought to SMARP. Specifically, the four types of datasets are generated as follows:
\begin{enumerate}
    \item Dataset \sharponly{}: 20\% of all the HARPs are randomly selected to form a test set. 20\% of the remaining HARPs are randomly selected to form a validation set. The rest of the HARPs belong to the training set. In each split, negative samples are randomly selected to match the number of positive samples.
    \item Dataset \fusedsharp{}: The test set and the validation set stay the same, respectively, with those in \sharponly{}. The remaining HARPs are combined with all TARPs to form the training set. In each split, negative samples are randomly selected to match the number of positive samples.
    \item Dataset \smarponly{}: 20\% of all the TARPs are randomly selected to form a test set. 20\% of the remaining TARPs are randomly selected to form a validation set. The rest of the TARPs belong to the training set. In each split, negative samples are randomly selected to match the number of positive samples.
    \item Dataset \fusedsmarp{}: The test set and the validation set stay the same, respectively, with those in \smarponly{}. The remaining TARPs are combined with all HARPs to form the training set. In each split, negative samples are randomly selected to match the number of positive samples.
\end{enumerate}
\begin{table}
    \centering
    \caption{Sample sizes of a random realization of the four datasets}
    \begin{tabular}{lcccccc}
        \toprule
        {} & \multicolumn{2}{c}{Train} & \multicolumn{2}{c}{Validation} & \multicolumn{2}{c}{Test} \\
        \cmidrule(lr){2-3} \cmidrule(lr){4-5} \cmidrule(lr){6-7}
        {} & Positive & Negative  &      Positive & Negative  & Positive & Negative  \\
        \midrule
        \texttt{SHARP\_ONLY}    &  1774 &  1774 &        665 &   665 &   410 &   410 \\
        \texttt{FUSED\_SHARP}   &  6375 &  6375 &        665 &   665 &   410 &   410 \\
        \texttt{SMARP\_ONLY}    &  2849 &  2849 &        860 &   860 &   892 &   892 \\
        \texttt{FUSED\_SMARP}   &  5698 &  5698 &        860 &   860 &   892 &   892 \\
        \bottomrule
    \end{tabular}
    \label{tab:data}
\end{table}
Since train/validation/test split and undersampling are both random, repeating these two steps with different seeds enables uncertainty quantification to the evaluation results.
The tally of samples produced by one particular random seed is shown in Table \ref{tab:data}. On each of the four types of datasets, LSTMs and CNNs are fitted on the training set, validated on the validation set, and evaluated on the test set.




\begin{table}
    \centering
    \caption{Test set performance of the LSTM and the CNN on 24-hour ``strong-vs-quiet" flare prediction. The two datasets within each comparison group share common test sets. The 1-$\sigma$ error is calculated from 10 random experiments. Bold fonts indicate the experiments in which the mean of the metric on the fused dataset is higher than that on the single dataset.}
    \input{table_member}
    \label{tab:member}
\end{table}

Table \ref{tab:member} shows the results of the ``strong-vs-quiet" active region prediction using the LSTM and the CNN.
For LSTMs, a consistent improvement on the fused datasets (\texttt{FUSED\_SHARP} and \texttt{FUSED\_SMARP}) is observed in terms of the mean of all metrics. This aligns with the fact that more data are typically desired to improve the generalization performance of deep learning models because they are overparameterized and can easily overfit on small datasets. For CNNs, an improvement is observed on \texttt{FUSED\_SMARP} over \texttt{SMARP\_ONLY}, but not on \texttt{FUSED\_SHARP} over \texttt{SHARP\_ONLY}. This indicates that the lower image quality in SMARP has a negative effect on CNN's performance.

The statistical significance of the improvement on the fused datasets is tested using a one-sided paired $t$-test with significance level 95\%. Table \ref{tab:ttest} shows the $t$-statistics and the associated $p$-values of the paired $t$-tests. The bold font $p$-values are less than 0.05 and considered to be significant. For LSTMs, the fused datasets are better than the single datasets in a statistically significant way in almost all settings. The only exception is BSS on \fusedsharp{}, whose $p$-value is only slightly larger than 0.05. For CNNs, across all metrics, statistically significant improvement is observed for \fusedsmarp{} over \smarponly{}, but not for \fusedsharp{} over \sharponly{}. This indicates that adding SHARP magnetograms into SMARP during training helps the CNN to better predict flares, but not the other way around. One potential reason is SMARP magnetograms have a lower signal-to-noise ratio than SHARP magnetograms, which may have negatively affected the CNN. The LSTM, on the other hand, uses the active region summary parameters, which could suppress the effect of noise during summarizing magnetograms, providing information in a sufficiently good quality that does not offset the improvement induced by the increased training sample size.

\begin{table}
    \centering
    \caption{Paired $t$-tests for significant improvement of test set performance on the fused datasets as measured by different metrics. The alternative hypothesis $H_1$ claims that metric $S$ on the fused dataset (\texttt{FUSED\_SHARP} or \texttt{FUSED\_SMARP}) is greater than the respective single dataset (\texttt{SHARP\_ONLY} or \texttt{SMARP\_ONLY}), which is tested against the null hypothesis $H_0$ claiming otherwise. The bold font $p$-values are less than 0.05 and considered to be significant.}
    \input{table_ttest}
    \label{tab:ttest}
\end{table}

\begin{figure}
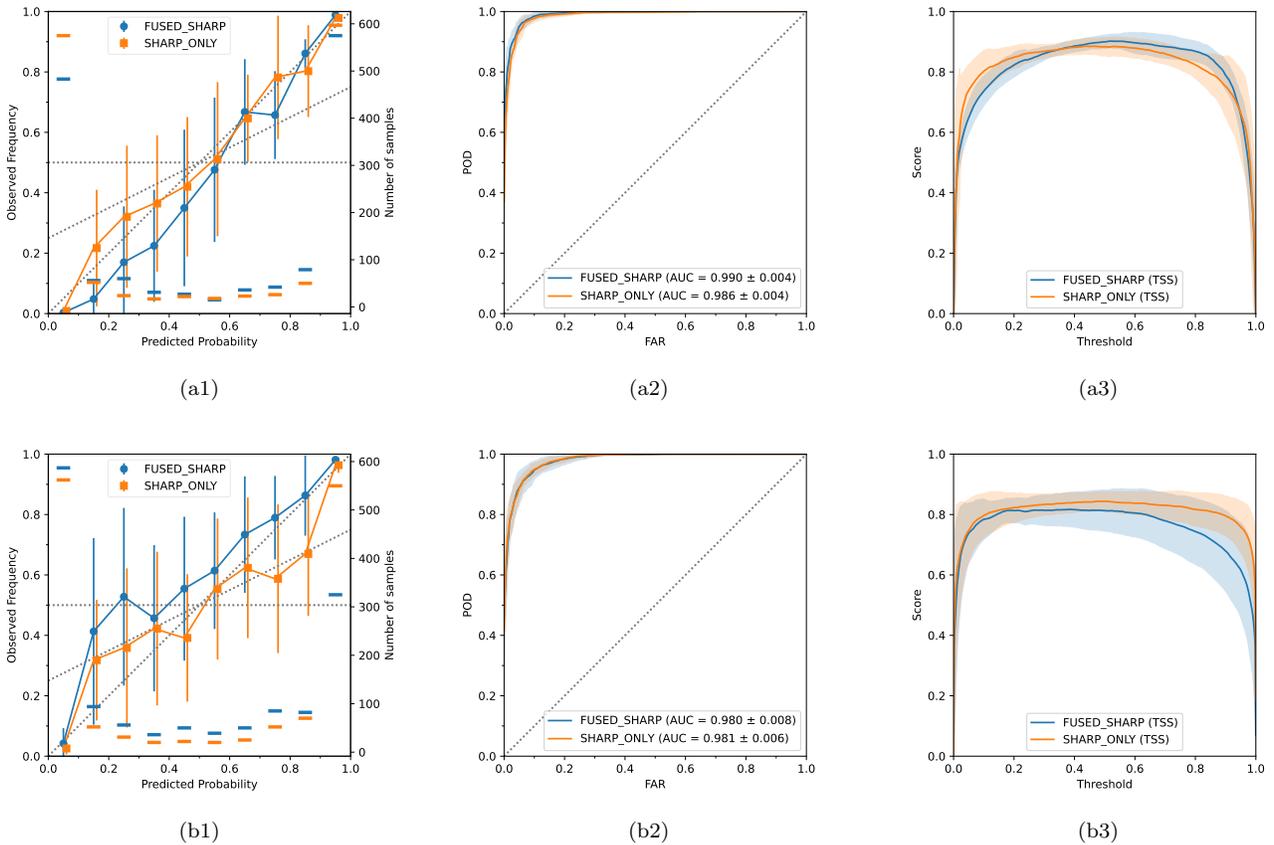

    \centering
    \gridline{
        \fig{graphical/sharp_LSTM/reliability}{0.3\textwidth}{(a1)}
        \fig{graphical/sharp_LSTM/roc}{0.3\textwidth}{(a2)}
        \fig{graphical/sharp_LSTM/ssp}{0.3\textwidth}{(a3)}
    }
    \gridline{
        \fig{graphical/sharp_CNN/reliability}{0.3\textwidth}{(b1)}
        \fig{graphical/sharp_CNN/roc}{0.3\textwidth}{(b2)}
        \fig{graphical/sharp_CNN/ssp}{0.3\textwidth}{(b3)}
    }
    \caption{Verification plots on SHARP test data to compare models trained on \fusedsharp{} and \sharponly{}. Shown in (a1)--(a3) are the reliability diagram, ROC, and SSP for LSTM. Shown in (b1)--(b3) are the same plots but for CNN. In each panel, the blue/orange curve is the test performance for the model trained on \fusedsharp{}/\sharponly{}. In each graph, solid curves and error bars (or shaded area) indicate respectively the means and the standard deviations calculated from 10 random experiments. In each reliability plot, the short horizontal bars indicate the number of samples in each probability bin, and the two curves are separated horizontally to prevent error bars from overlapping.}
    \label{fig:graphical_sharp}
\end{figure}

\begin{figure}
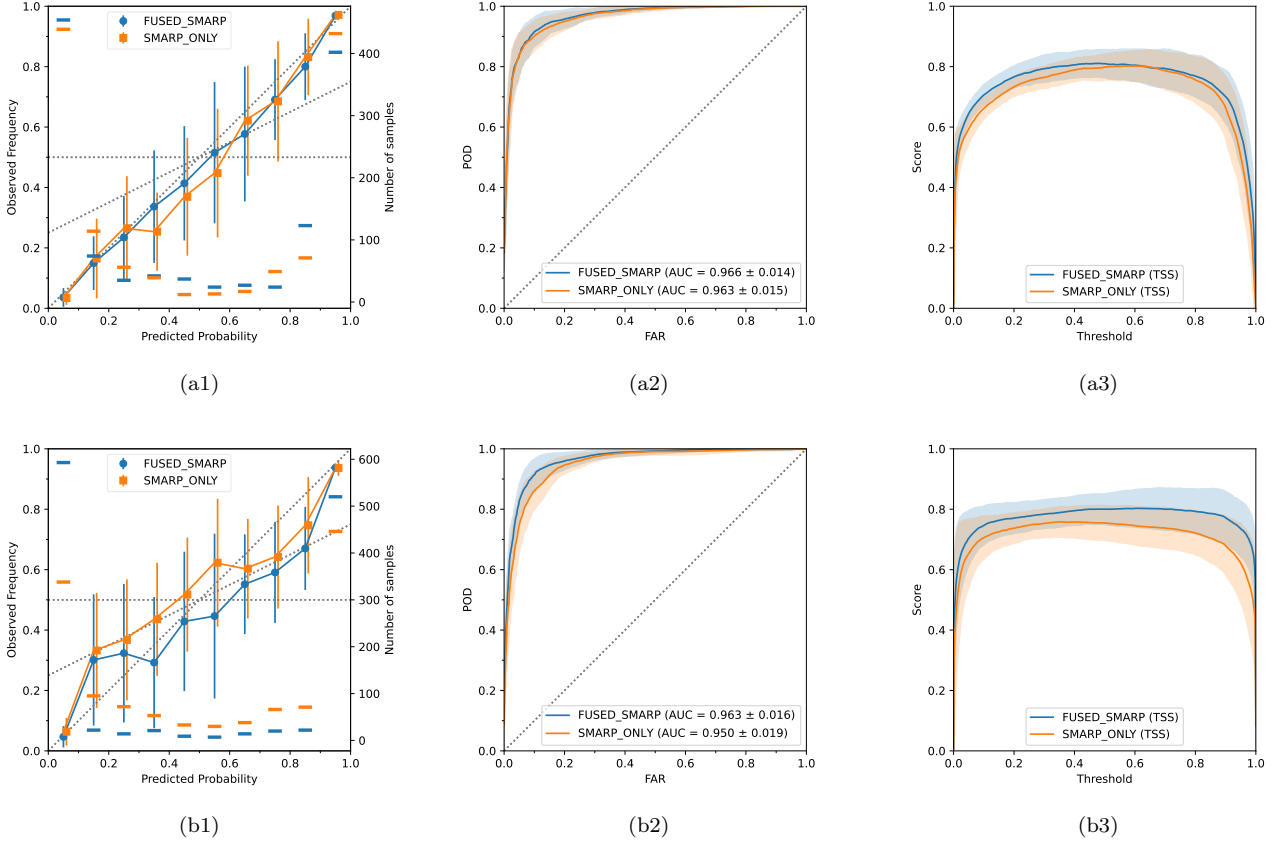

    \centering
    \gridline{
        \fig{graphical/smarp_LSTM/reliability}{0.3\textwidth}{(a1)}
        \fig{graphical/smarp_LSTM/roc}{0.3\textwidth}{(a2)}
        \fig{graphical/smarp_LSTM/ssp}{0.3\textwidth}{(a3)}
    }
    \gridline{
        \fig{graphical/smarp_CNN/reliability}{0.3\textwidth}{(b1)}
        \fig{graphical/smarp_CNN/roc}{0.3\textwidth}{(b2)}
        \fig{graphical/smarp_CNN/ssp}{0.3\textwidth}{(b3)}
    }
    \caption{Same as Figure \ref{fig:graphical_sharp} but for SMARP test data to compare models trained on \fusedsmarp{} and \smarponly{}.}
    \label{fig:graphical_smarp}
\end{figure}

Aside from the numerical metrics, we provide graphical evaluation results for Group 1 (\fusedsharp{} and \sharponly{}) in Figure \ref{fig:graphical_sharp}, and Group 2 (\fusedsmarp{} and \smarponly{}) in Figure \ref{fig:graphical_smarp}. A trend of over-forecasting for high probabilities and under-forecasting for low probabilities is observed in some cases but such effect is minor considering the size of the error bars.
In reliability diagrams, all models have points closer to the diagonal, indicating high reliability.
In ROC plots, it is observed that the LSTM achieves higher AUC on the fused datasets (\texttt{FUSED\_SHARP} and \texttt{FUSED\_SMARP}) than on the single datasets (\texttt{SHARP\_ONLY} and \texttt{SMARP\_ONLY}). For the CNN, similar improvement is also observed in the comparison of \fusedsmarp{} and \fusedsharp{}, whereas the ROCs are almost indistinguishable for \fusedsharp{} and \sharponly{}.
In skill score profiles, the TSS for LSTM trained on fused datasets are at the same level as that trained on single datasets. For the CNN, on the other hand, \fusedsharp{} displays a disadvantage against \sharponly{}, whereas \fusedsmarp{} displays an advantage over \smarponly{}. This verifies the observations made from metrics. In all cases, the skill score profiles are high and relatively flat, indicating the robustness of the performance to the change of thresholds within a wide range of the varying threshold.

\subsection{LSTM performs better than CNN}
\label{sec:results_model}



This section provides forecast verification to the LSTM and the CNN. We use the same evaluation results for 10 experiments in each setting mentioned in Section \ref{sec:results_data}, but present them in a way that makes it easier to compare the LSTM and the CNN. We note the differences between our verification set-up and that in an operational setting:
\begin{enumerate}
    \item In terms of data, the test set of our sort has lots of samples removed based on their active regions, observational data, and flare activities. About 1/5 of tracked active region time series in the evaluation period (May 2010--December 2020) are selected. Within each active region series, only samples with good quality observation and certain flaring patterns are selected (detailed in Section \ref{sec:sample}). Negative samples (flare-quiet active regions) are significantly downsampled to match the number of positive samples (strong-flare-imminent active regions). In contrast, operational forecasts do not discard any sample unless absolutely necessary.
    \item In terms of outcomes, the forecast of our sort is independent for individual active regions, with the prediction result available every 96 minutes (i.e., MDI cadence) for valid active regions. In contrast, the end goal of an operational forecast is a full-disk forecast. For operational forecasts built upon active region based forecasts, the predictions for all active regions on the solar disk are aggregated to compute the full-disk prediction. In addition, operational forecasts are typically issued at a lower frequency (e.g., every 6 hours), but in a consistent manner.
\end{enumerate}
The verification results in this section should be interpreted with the above differences in mind.



\begin{table}
    \centering
    \caption{Paired $t$-tests for significant improvement of the LSTM over the CNN in terms of different metrics $S$ on the test set of the four datasets. The alternative hypothesis $H_1$ claims $S_{\textrm{LSTM}} > S_{\textrm{CNN}}$. The bold font $p$-values are less than 0.01 and considered to be significant.}
    \input{table_ttest_est}
    \label{tab:ttest_est}
\end{table}

It can be seen from Table \ref{tab:member} that the LSTM generally scores higher than the CNN in terms of mean performance. We performed paired $t$-test to validate this observation. The results in Table \ref{tab:ttest_est} confirm that the LSTM scores significantly higher ($p<0.01$) than the CNN across all metrics on all datasets except for \fusedsmarp. On \fusedsmarp, although we cannot claim statistical significance, the LSTM's performance is slightly better or at the same level as the CNN as is observed from Table \ref{tab:member}.

We only present the graphical verification results for both models trained and tested \fusedsharp{}, given that SHARP is widely used and validated by a wealth of studies. For the results on other datasets, the visualization can be obtained by simply rearranging the same results shown in Figure \ref{fig:graphical_sharp} and \ref{fig:graphical_smarp}.

\begin{figure}
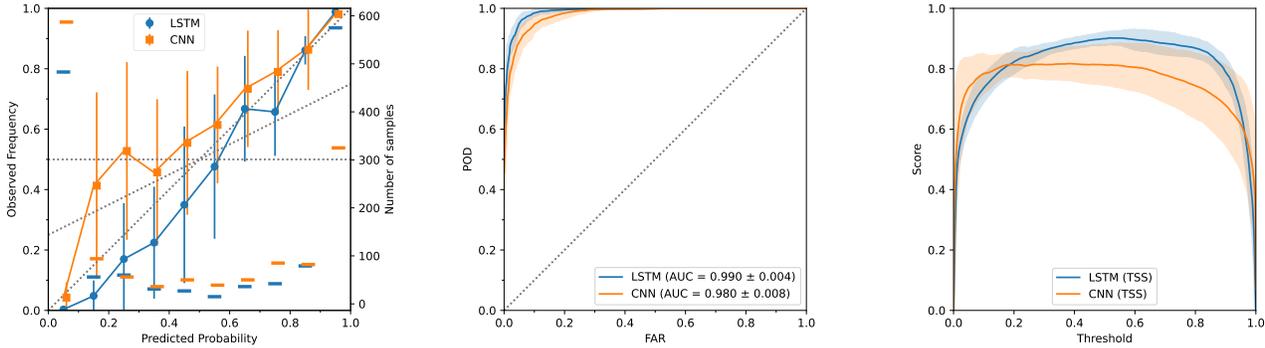

    \centering
    \gridline{
        \fig{graphical/model_comp_fused_sharp/reliability}{0.3\textwidth}{}
        \fig{graphical/model_comp_fused_sharp/roc}{0.3\textwidth}{}
        \fig{graphical/model_comp_fused_sharp/ssp}{0.3\textwidth}{}
    }
    \caption{Verification plots of the LSTM and the CNN on \fusedsharp{}. Shown are the reliability diagram, ROC, and SSP, from left to right. This figure essentially extracts the blue curves (representing \fusedsharp{}) in both rows of Figure \ref{fig:graphical_sharp} and overlaps them together.}
    \label{fig:graphical_model_comp}
\end{figure}

The reliability diagram in Figure \ref{fig:graphical_model_comp} shows that the probabilistic prediction given by the LSTM is closer to the diagonal than the CNN, and hence more reliable. The CNN exhibits a trend of under-forecasting especially when the predicted probability is less than 0.5. The histogram of predicted probability shows that probabilistic forecast by the LSTM is ``more confident", or has higher resolution, than LSTM, with most of the predicted probabilities close to 0 or 1. 

The ROC in Figure \ref{fig:graphical_model_comp} shows a clear advantage of the LSTM over the CNN, in the sense that it achieves a higher probability of detection with the same false alarm rate. This trend is also manifested in terms of AUC.

The SSP in Figure \ref{fig:graphical_model_comp} shows LSTM achieves higher TSS on average for all thresholds within 0.2--0.9. It is also observed that the TSS for LSTM is maximized by a threshold very close to the climatological rate on the test set (which is 0.5 in our case), a necessary condition for a reliable predictor \citep{kubo2019some}.

\subsection{Stacking LSTM and CNN leads to better prediction}
\label{sec:results_stacking}


In this paper, we only consider stacking methods to combine the CNN and the LSTM hoping for better predictive performance. We evaluate the test set performance of stacking methods using four different criteria:
\begin{itemize}
    \item \texttt{CROSS\_ENTROPY}: weights are optimized to minimize cross-entropy loss on the validation set.
    \item \texttt{BSS}: weights are optimized to maximize BSS on the validation set.
    \item \texttt{AUC}: weights are optimized to maximize AUC on the validation set.
    \item \texttt{TSS}: weights are optimized to maximize TSS on the validation set.
\end{itemize}
Among these criteria, cross-entropy and negative $\BSS$ are known to be convex; $\TSS$ is neither convex nor concave; we observe $\AUC$ to be concave but we do not have proof other than empirical evidence. Criteria $\texttt{HSS}$ and $\texttt{ACC}$ are excluded from the evaluation since their stacking weights are the same as that of $\texttt{TSS}$ due to the perfect correlation mentioned in Section \ref{sec:evaluation}.

To provide baseline performances, we include the evaluation results for the two base learners, \texttt{LSTM} and \texttt{CNN}. In addition to the above stacking methods, we consider two other meta-learning schemes:
\begin{itemize}
    \item \texttt{AVG} outputs the average of predicted probabilities of two base learners.
    \item \texttt{BEST} \citep{dvzeroski2004combining} selects the base learner that performs the best on the validation set and applies it to the test set.
\end{itemize}

\begin{figure}
    \centering
    \includegraphics[width=\textwidth]{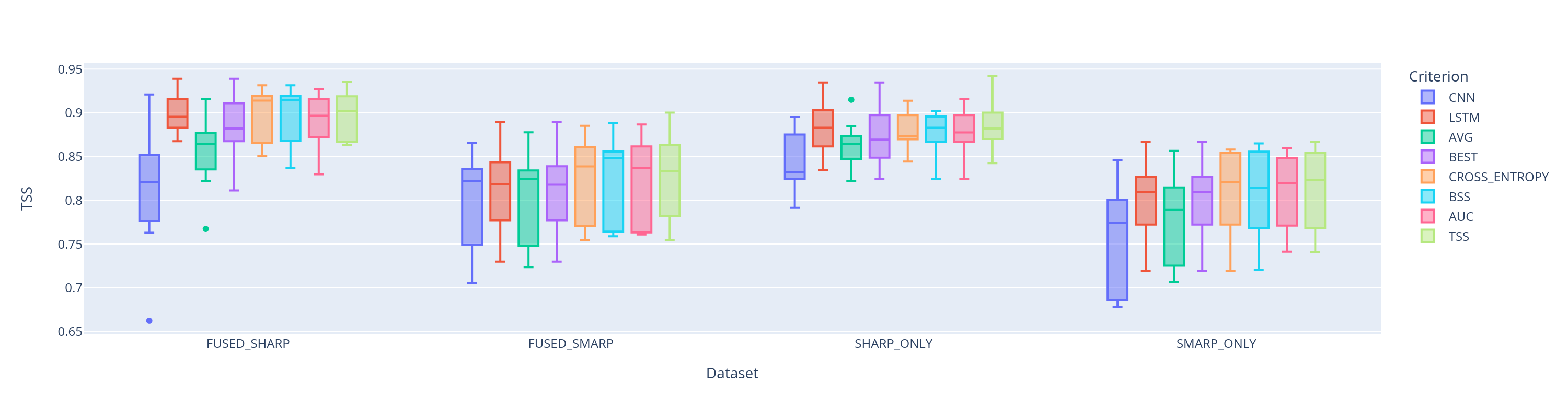}
    \caption{Test set TSS for base learners and meta-learners using different criteria.}
    \label{fig:stacking}
\end{figure}

\begin{figure}
    \centering
    \includegraphics[width=\textwidth]{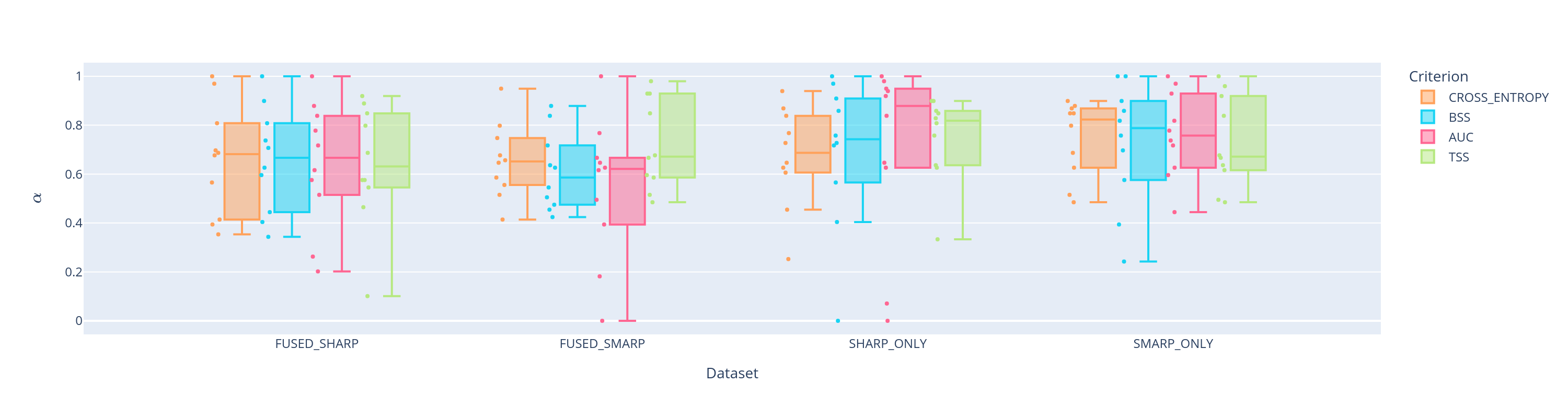}
    \caption{Stacking weight $\alpha$ fitted using different criteria on different datasets. All 10 values of $\alpha$ in an experiment setting are shown as points next to the corresponding box.}
    \label{fig:stacking_weights}
\end{figure}

Splitting and undersampling are randomly performed 10 times on each of the four datasets \fusedsharp{}, \fusedsmarp{}, \sharponly{}, and \smarponly{}. The test set TSS of the 10 random experiments for each criterion on each dataset are summarized as box plots in Figure \ref{fig:stacking}. The optimal stacking weights for the four stacking ensembles are summarized in Figure \ref{fig:stacking_weights}.



Figure \ref{fig:stacking} shows that stacking methods perform slightly better than the \texttt{BEST} meta-learner, especially on \fusedsmarp{} and \smarponly{}. Of note, the wide error bars are partially due to the randomness originating from data sampling. To fairly compare the methods, we perform paired $t$-tests with significance level 0.05. It turned out stacking is significantly better than \texttt{BEST} in the following three settings: \texttt{BSS} on \fusedsmarp{} ($p = 0.048$), \texttt{AUC} on \smarponly{} ($p = 0.025$), and \texttt{TSS} on \smarponly{} ($p = 0.013$).

We also note in Figure \ref{fig:stacking} that \texttt{BEST} unsurprisingly achieves better performance than \texttt{AVG} but is slightly worse than the better performing base learner \texttt{LSTM}, most noticeably on \fusedsharp{}. In fact, \texttt{BEST} decided that \texttt{CNN} is the better model in 3 out of 10 experiments on \fusedsharp{}. This is not unexpected because the ``best" model on the validation set is not necessarily the best on the test set.

From Figure \ref{fig:stacking_weights}, we can see that $\alpha$ is greater than 0.5 in 
most 
experiments, with the median falling between 0.55 and 0.9 in all settings. This suggests that stacking ensembles generally depend more on the LSTM than on the CNN. The variance of $\alpha$ is large in some settings, especially for the \texttt{AUC} on \fusedsmarp{}.
The variance of convex criteria (\texttt{CROSS\_ENTROPY} and \texttt{BSS}) is not smaller than that of nonconvex criteria (\texttt{TSS}), indicating that the local minima of non-convex loss functions is not the major source of variance. We suspect the major source of the variance comes from the data sampling bias among experiments, which is, in turn, a collective consequence of the insufficient sample size, heterogeneity across active regions, and possibly a small amount of information leakage because the validation set is used both in the validation of base learners and the training of the meta-learner.

\begin{figure}
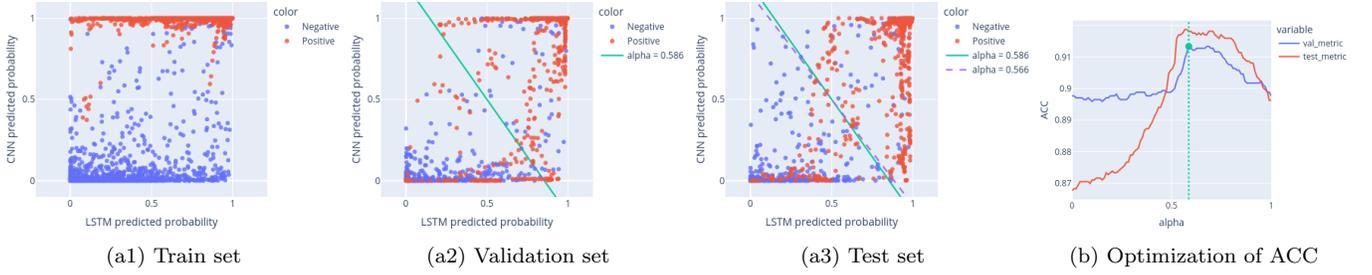

    \centering
    \gridline{
        \fig{stacking/data_train.png}{0.25\textwidth}{(a1) Train set}
        \fig{stacking/data_val.png}{0.25\textwidth}{(a2) Validation set}
        \fig{stacking/data_test.png}{0.25\textwidth}{(a3) Test set}
        \fig{stacking/alpha.png}{0.23\textwidth}{(b) Optimization of ACC}
    }
    \caption{(a1)--(a3): CNN predicted probability (y-axis) vs. LSTM predicted probability (x-axis) for the train, the validation, and the test set. The green solid line in (a2) and (a3) is the decision boundary of the ensemble with meta-learner fitted on the validation set. The violet dashed line in (a3) is analogous to the green line except that it is fitted on the test set, and hence can be seen as the oracle. (b): ACC as a function of $\alpha$ on the validation and the test set. The vertical green line shows the value of $\alpha$ that maximizes the validation ACC. The leftmost values of the ACC curves ($\alpha=0$) correspond to the ACC of the CNN, and the rightmost values of these curves ($\alpha=1$) correspond to the ACC of the LSTM.}
    \label{fig:data}
\end{figure}

We inspect one experiment of stacking with criterion \texttt{ACC} and the results are presented in Figure \ref{fig:data}. Figure \ref{fig:data} (a1)--(a3) show the predicted probabilities by the LSTM and the CNN of each instance in the training, the validation, and the test set. The points are colored by their labels, with red representing the positive class and blue representing the negative class. The green solid line in (a2) and (a3) shows the decision boundary by the meta-learner with $\alpha$ fitted on the validation set to maximize ACC. The points $(p, q)$ on the upper right side of the boundary are classified as positive because they satisfy $r = \alpha p + (1-\alpha) q > 0.5$. In this experiment, the fitted $\alpha=0.586$, suggesting the stacking ensemble relies almost equally on the CNN and the LSTM. The violet dashed line in (a3) is the decision boundary with $\alpha$ fitted on the test set, and hence can be seen as the oracle. It can be observed that the distribution of predicted probabilities on the validation set (a2) and the test set (a3) are similar. The distribution of predicted probabilities on the training data in (a1), on the other hand, looks completely different, with the CNN achieving almost perfect separation. In fact, the CNN overfitted on the in-sample data, as indicated by a significantly lower positive recall rate in (a2) and (a3). This validates the decision that meta-learners should not be fitted on the predicted probabilities of the same data used to train the base learners.

Figure \ref{fig:data}(b) exhibits the stacking optimization process for the same experiment, in which the ACC is calculated on the validation set (a2) and the test set (a3) by scanning over a fine grid of $\alpha\in[0,1]$ with resolution 0.001. Although the validation ACC (blue curve) is not concave with respect to $\alpha$, it does exhibits a maximum at $\alpha=0.586$ as indicated by the vertical green line. The stacking ensemble's test set performance over $\alpha$ is shown as the red curve. These curves indicate that stacking the LSTM and the CNN indeed results in a small improvement of performance relative to implementing either of them alone, corresponding to the values of ACC at $\alpha=1$ or $\alpha=0$.

\subsection{CNN identifies the emergence of preflare features}


We use visual attribution methods to extract flare-indicative characteristics of magnetograms from trained CNNs.
First, we use synthetic images to examine patterns that contribute to a positive decision of CNNs. The results of synthetic images help us understand better the attribution maps of real magnetograms.
Then, we apply visual attribution methods to image sequences of selected active regions that transition from a flare-quiescent state to a flare-imminent state. Setting the baseline to the first image in the sequence gives a time-varying attribution map that tracks magnetic field variations that contribute to the change in the predicted probability.

\subsubsection{Synthetic image}

To assist our understanding of attribution maps obtained by different methods, we first turned to synthetic magnetograms. We take the bipolar magnetic region (BMR) model in \citet{yeates2020good}, represented as line-of-sight magnetic field $B$ as a function of Heliographical location $(s,\phi)$, where $s$ denotes sine-latitude and $\phi$ denotes Carrington longitude. $B$ is parameterized by amplitude $B_0$, polarity separation $\rho$ (in radian), tilt angle $\gamma$ (in radian) with respect to the equator, and size factor $a$ fixed to be 0.56 to match the axial dipole moment of SHARP \citep{yeates2020good}. The untilted BMR centered at origin has the form
\begin{align}
    B(s, \phi) = - B_0 \frac{\phi}{\rho} \exp\left[-\frac{\phi^2 + 2 \arcsin^2(s)}{(a\rho)^2}\right].
\end{align}

\begin{figure}
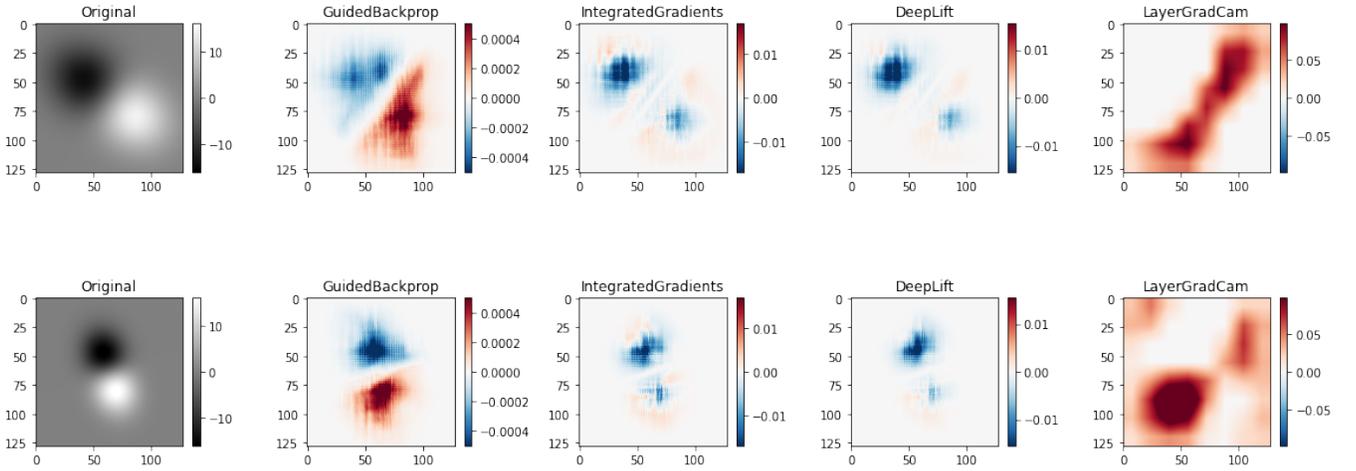

    \centering
    \gridline{\fig{syn/0.png}{\textwidth}{}}
    \gridline{\fig{syn/5.png}{\textwidth}{}}
    \caption{Examples of synthetic bipole images and attribution maps.}
    \label{fig:bipole}
\end{figure}

We sweep a grid of $B_0$, $\rho$, and tilt angle $\gamma$ to generate a BMR dataset. Of particular interest are synthetic BMRs considered to be flare-imminent by CNNs. Figure \ref{fig:bipole} shows some examples of them and their attribution results, from which patterns of positive predictions can be summarized. Guided Backpropagation heatmaps have both poles highlighted with the signs matching the polarities. Integrated Gradients produces heatmaps that are more concentrated to polarity centers and attribute more credits to the negative polarities. DeepLIFT produces similar heatmaps to those by Integrated Gradients. Grad-CAM's results are not as interpretable as the above methods. They seem to avoid the polarities and highlight the background and sometimes the polarity inversion lines.

\subsubsection{The emergence of preflare signatures in the active region evolution}

\begin{figure}
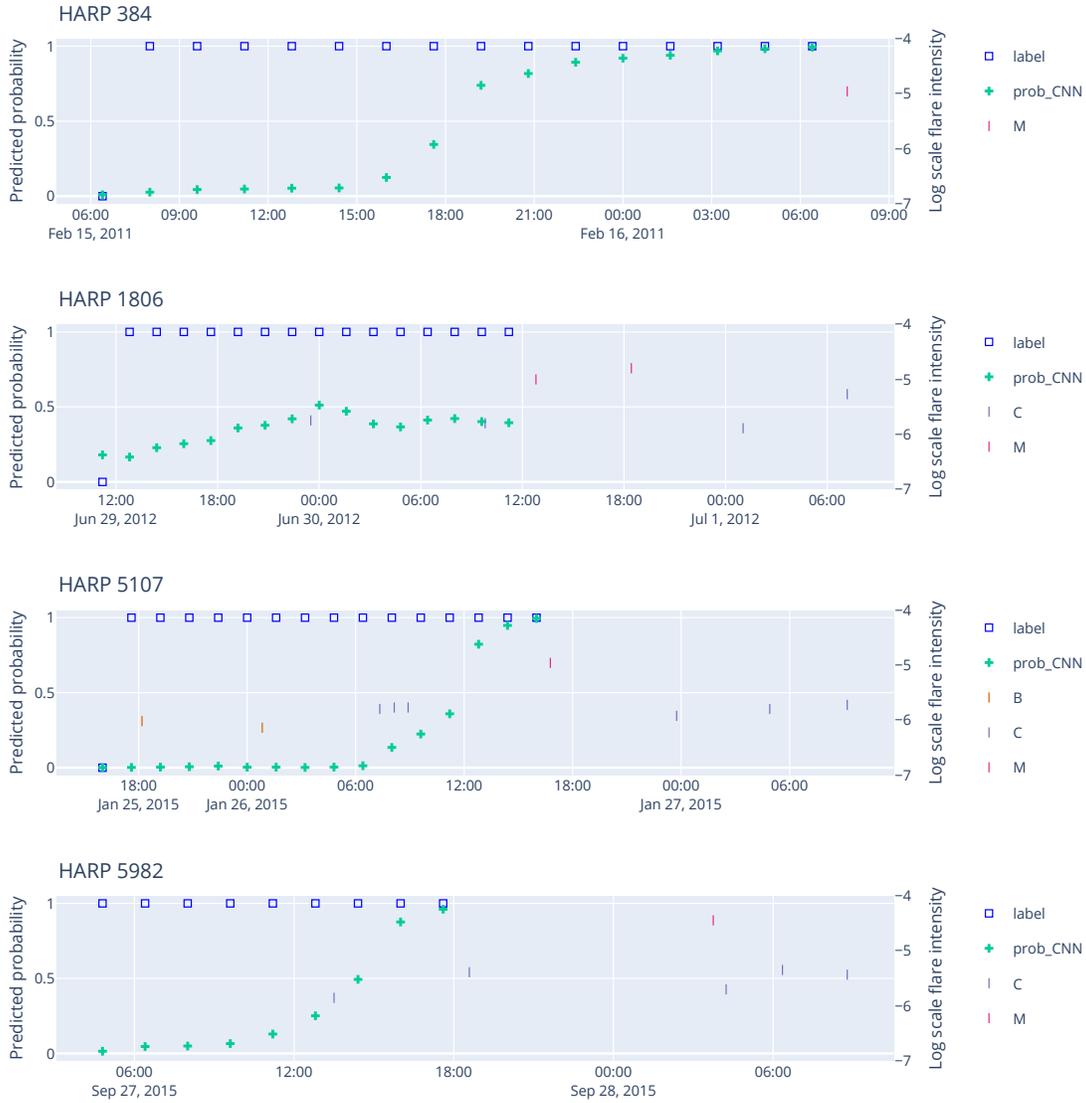

    \centering
    \gridline{\fig{contour/384/time_series.pdf}{0.8\textwidth}{}}\vspace{-10px}
    \gridline{\fig{contour/1806/time_series.pdf}{0.8\textwidth}{}}\vspace{-10px}
    \gridline{\fig{contour/5107/time_series.pdf}{0.8\textwidth}{}}\vspace{-10px}
    \gridline{\fig{contour/5982/time_series.pdf}{0.8\textwidth}{}}\vspace{-10px}
    \caption{CNN predictions of part of time series of in HARP 384, 1806, 5107, and 5982. The labels are shown as blue open boxes and predicted probabilities as green plus symbols. The point-in-time instance is labeled as positive if an M1.0+ flare occurred in the future 24 hours in that active region. GOES flare events during and 24 hours within the sample sequence are shown as short vertical bars, with y-coordinates indicating flare intensities (peak flux in W/m$^2$) on a log scale. 
    }
    \label{fig:time_series}
\end{figure}

We focus on the attribution results on SHARP as opposed to SMARP because the former has magnetograms of higher resolution and lower noise level. We choose the CNNs that are trained on \sharponly{} as opposed to \fusedsharp{} because the former is observed to generalize better according to Section \ref{sec:data}. To get results that reflect the generalization performance as opposed to training artifacts, we need to make sure that active regions being investigated are out-of-sample. To evaluate any active region of interest in SHARP, we perform 5-fold cross-validation on \sharponly{}, so that every active region is associated with a CNN that has never seen the active region in training. In addition, we do not enforce the flare-based sample selection rules and random undersampling, so that the evolution of attribution maps can be evaluated more coherently.
As case studies, we select four HARP sequences that transition from a flare-quiescent state to a flare-imminent state.
Figure \ref{fig:time_series} shows the labels and predicted probabilities of the four sample sequences.
The attribution methods are performed on each HARP sequence in a frame-by-frame manner.

\begin{figure}
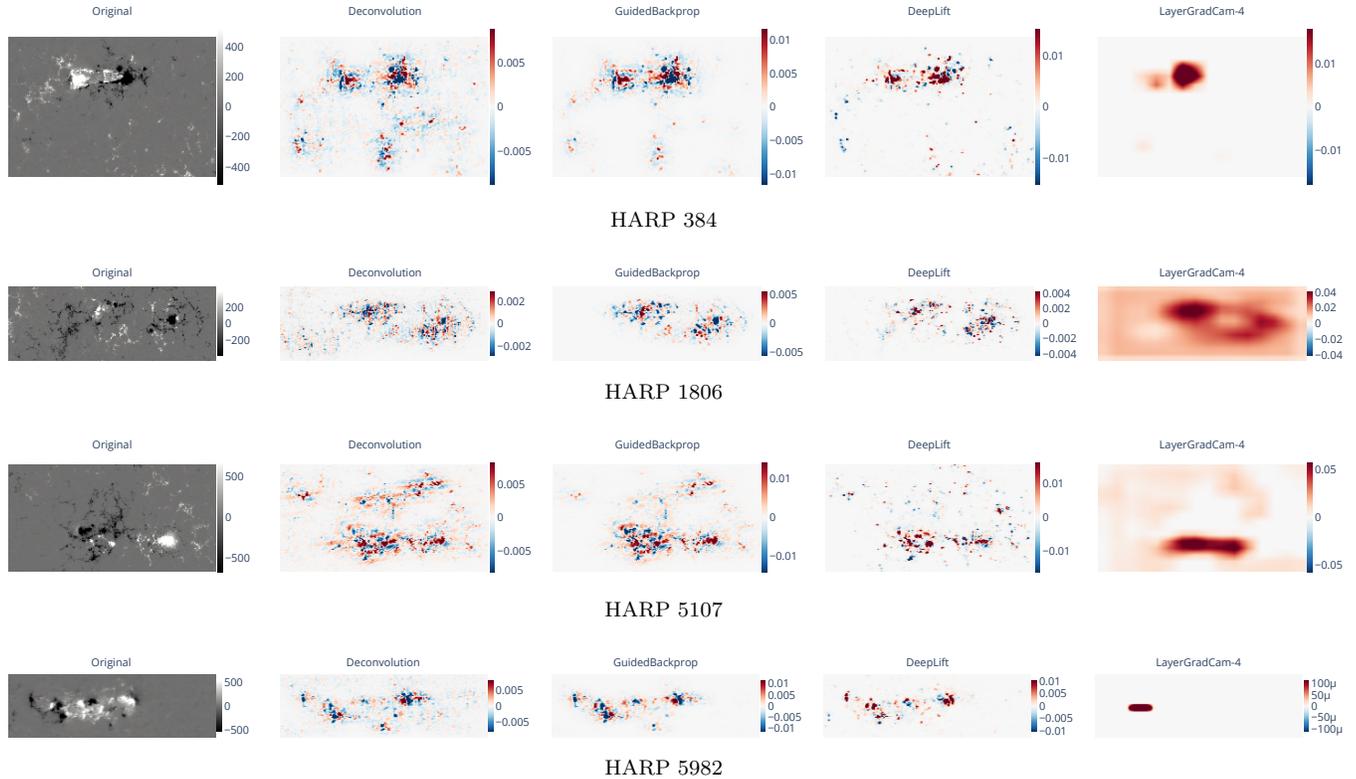

    \centering
    %
    \gridline{\fig{contour/384/last.pdf}{\textwidth}{HARP 384}}
    \gridline{\fig{contour/1806/last.pdf}{\textwidth}{HARP 1806}}
    \gridline{\fig{contour/5107/last.pdf}{\textwidth}{HARP 5107}}
    \gridline{\fig{contour/5982/last.pdf}{\textwidth}{HARP 5982}}
    \caption{Attribution results of Deconvolution, Guided Backpropagation, DeepLIFT, and Grad-CAM on the last magnetogram in the sample sequences of HARP 384, 1806, 5107, and 5982. DeepLIFT chooses the first sample in the sequence as the reference. ``LayerGradCam-4" means Grad-CAM with respect to the output of the fourth, or the second to last, convolutional layer. The interactive movie of heatmaps on all 9 samples in HARP 5982 using more attribution methods can be accessed at \url{https://zeyusun.github.io/attribution/captum_movie_first.html}.}
    \label{fig:attribution}
\end{figure}

Figure \ref{fig:attribution} shows the last image of the four HARP sample sequences. The attribution maps of the same size as the input of the CNN (128$\times$ 128 pixels) are upsampled to the original resolution of the SHARP magnetogram using the \texttt{resize} method of the Python package \texttt{skimage.transform} with 2nd-order spline interpolation. The attribution maps of DeepLIFT and Integrated Gradients are similar. As such, only the results of the former are shown. The results for Integrated Gradients can be accessed online with the link shown in the caption.

In Figure \ref{fig:attribution}, the attribution maps of Guided Backpropagation are observed to be more concentrated in strong fields compared to that of Deconvolution. The reference image of DeepLIFT and Integrated Gradients are chosen as the first sample in each sequence. From these two methods, the change of the prediction scores is attributed to the change of magnetic configuration of the last frame relative to the first frame, with red pixels indicating positive contribution and blue pixels indicating negative contribution. Since the predicted event probability of the last frame is higher than the first frame for all HARPs (Figure \ref{fig:time_series}), the red pixels outweigh the blue pixels in the attribution maps of DeepLIFT and Integrated Gradients. The Grad-CAM results roughly reveal the position of the strong fields and polarity inversion lines.

\begin{figure}
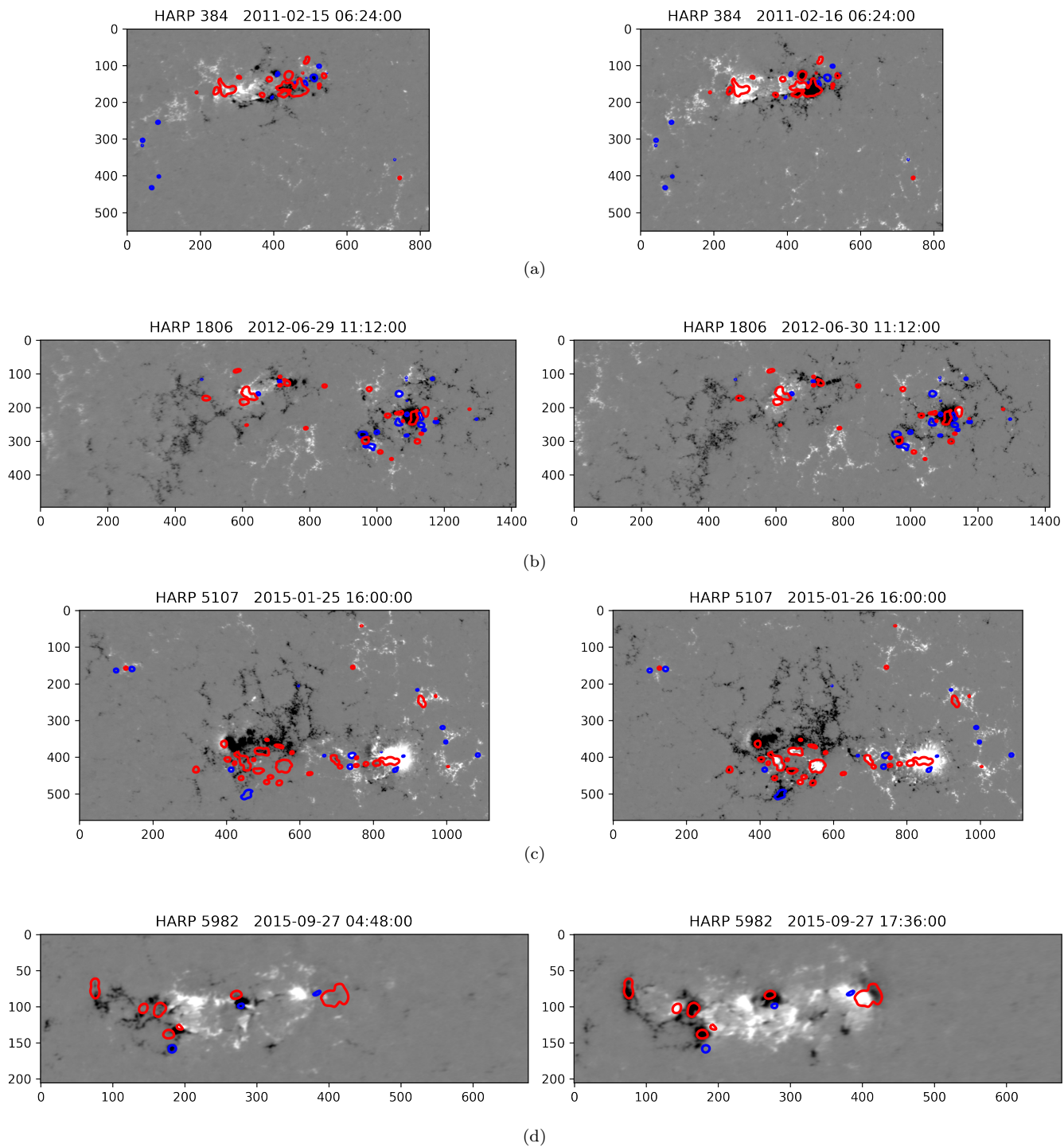

    \centering
    \gridline{\fig{contour/384/15.png}{\textwidth}{\vspace{-15px}(a)}}\vspace{-15px}
    \gridline{\fig{contour/1806/15.png}{\textwidth}{\vspace{-15px}(b)}}\vspace{-15px}
    \gridline{\fig{contour/5107/15.png}{\textwidth}{\vspace{-15px}(c)}}\vspace{-15px}
    \gridline{\fig{contour/5982/8.png}{\textwidth}{\vspace{-20px}(d)}}\vspace{-15px}
    \caption{Highly attributed pixels in the last frame by Integrated Gradients on four select HARPs shown in rows. In (a), the left/right panel shows the first/last magnetogram in the sample sequence of HARP 384. The magnetograms are in the SHARP resolution, with ticks on the axes indicating pixels. Pixel values saturate at $\pm 500$~Gs. The red/blue contours on the right panel (last frame) highlight the areas with strong positive/negative Integrated Gradients relative to the first frame. The same contours are mapped to the left panel (first frame) for contrast.
    The contours are drawn on the attribution map smoothed with a Gaussian kernel with a standard deviation of 3 pixels.
    Figures in (b), (c), and (d) are similar to (a) but for other HARPs.
    An animated version of this figure that serially presents the complete sample sequences of the four HARPs is available online.}
    \label{fig:contour}
\end{figure}

From the visual attribution map, the CNN's prediction of a flaring active region can be accredited to the elements in the magnetogram. Figure \ref{fig:contour} shows the contour plots of attribution maps generated by Integrated Gradients overlaid on magnetograms of the four HARP series. The contours enclose areas with large absolute values of Integrated Gradients in the last frame of each series, with red/blue contours indicating the region contributing positively/negatively to the increase in predicted probability. A general pattern is that the flux is emerging in red contours and canceling in blue contours. From the attribution maps, we can explain the increase in prediction scores as the consequence of the emerging flux outweighing the canceling flux. 

\begin{figure}
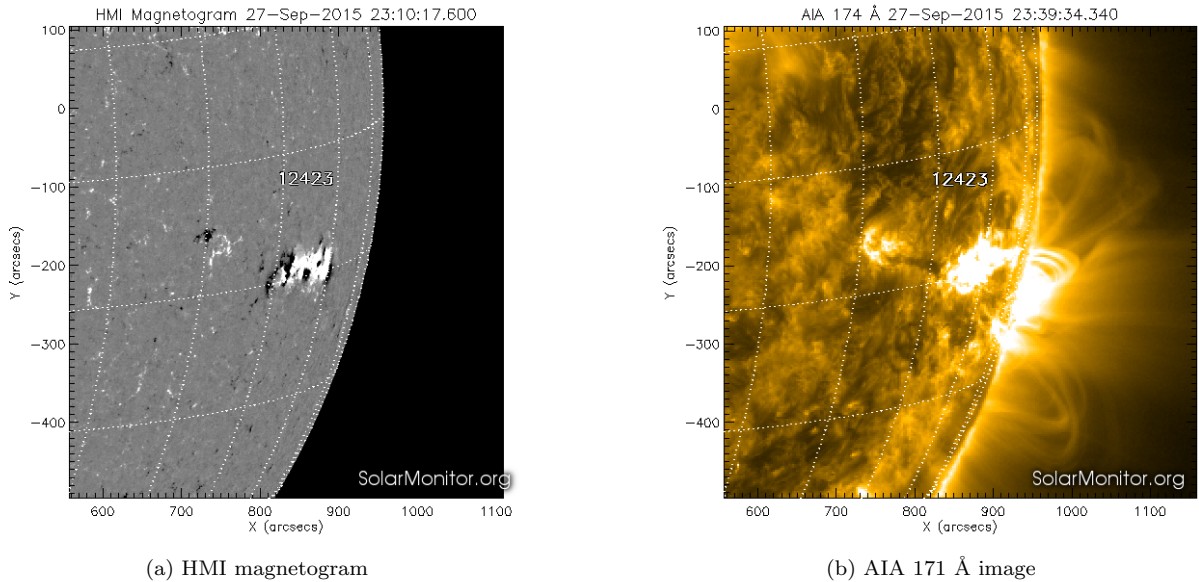

    \centering
    \gridline{
        \fig{contour/5982/hmi_ar_12423.png}{0.4\textwidth}{(a) HMI magnetogram}
        \fig{contour/5982/aia_171_ar_12423.png}{0.4\textwidth}{(b) AIA 171 \AA{} image}
    }
    \caption{Line-of-sight magnetic field (a) and solar EUV image (b) of HARP 5982 (NOAA AR 12423) at 23:10:17 on Sep 27, 2015. Images are taken from \url{https://solarmonitor.org/}. Note that the image title of (b) should be ``AIA 171~\AA{}" instead of ``AIA 174~\AA{}".}
    \label{fig:harp5982}
\end{figure}

The visual attribution maps can not only be used to identify preflare signatures in an active region; comparing them with our knowledge of flaring active regions can provide insights to diagnose, and potentially improve, the machine learning method used to predict flares. Here we provide an example. A known artifact in magnetograms is the fake polarity inversion line (PIL) caused by the projection effect when the magnetic vector's inclination relative to the line-of-sight surpasses 90$^\circ$ \citep{leka2017evaluating}. In Figure \ref{fig:contour}(d), the emerging polarity inversion line in the penumbra of the leading polarity (on the right/west part of the active region) is picked up as a preflare signature by the largest red contour. However, HARP 5982 is on the limb of the solar disk at the time (Figure \ref{fig:harp5982}), and the emerging PIL is caused by the highly inclined magnetic field in the penumbra as the flux rope is elevating from the surface. This shows that the CNN trained to associate magnetograms and flaring activities is not able to discern the polarity artifact by itself. This also suggests that the model could be potentially improved if we feed the location information to the CNN to help it correct such artifact.
A similar PIL artifact is also observed in the following polarity of HARP 5107 in Figure \ref{fig:contour}(c). Since this artifact does not change much during the observation interval, it does not contribute as much to the change of the prediction score.

We remark the attribution maps obtained by Integrated Gradients are better in terms of resolution and interpretability than what were used in \citet{bhattacharjee2020supervised} and \citet{yi2021visual}.
The occlusion method in \citet{bhattacharjee2020supervised} was shown to highlight the area between the opposite polarities, providing only crude attribution. This is because the size of the occlusion mask is usually chosen to be big enough to cover the informative regions.
The result of Grad-CAM, being the attribution to a convolutional layer as opposed to the input, also suffers from the low-resolution issue. Both the Grad-CAM results in Figure \ref{fig:attribution} and in \cite{yi2021visual} are able to highlight active regions, but the resolution is not high enough to reveal any structural information within the active region at the level of magnetic elements.
Guided Backpropagation in \citet{yi2021visual} is able to identify polarity inversion lines. However, it has been observed (and theoretically assessed) that Guided Backpropagation and Deconvolution behave similarly to an edge detector, i.e., they are activated by strong gradients in the image and insensitive to network decisions \citep[e.g.][]{nie2018theoretical,adebayo2018sanity}.
In contrast, the method of Integrated Gradients needs a baseline, which aligns with the natural way in which the human interprets an observation: by assigning credit or blame to a certain cause, we implicitly consider the absence of the cause \citep{sundararajan2017axiomatic}. In addition, Integrated Gradients essentially ``decomposes" the change of the network's prediction score to pixels in the input image, leading to a high-resolution attribution map.


\section{Conclusions and discussion}
\label{sec:discussion}

In this paper, we used two solar cycles of active region observational data from SMARP \citep{bobra2021smarps} and SHARP to examine the improvement in flare predictive performance of two deep learning models, namely the LSTM and the CNN, when trained on the fused datasets. When tested on SMARP, both models showed significant improvement. When tested on SHARP, LSTM showed significant improvement. The results of the controlled comparative studies indicate such an improvement is due to the significantly increased sample size from the other solar cycle. Then, in our setting of flare prediction, we verified the performance of the LSTM and the CNN using skill scores, reliability diagrams, ROC, and skill score profiles. The comparison showed that the LSTM is generally a better model than the CNN. After that, we explored the possibility of combining the LSTM and the CNN for a better prediction performance in the framework of a meta-learning paradigm called stacking. The results showed that in some settings, the stacking model outperforms the best member in the ensemble. Lastly, we applied visual attribution methods to CNNs. The results demonstrate the utility of visual attribution methods in identifying flare-related signatures in active regions, including the flux emergence and new polarity inversion lines. The attribution map on one particular region on the limb of the solar disk revealed one limitation of the CNN and suggested potential modifications for improvement.

The questions raised in Section \ref{sec:results} are arguably broad and general. We have taken one particular path to partially address each question. To inspire future studies, we provide additional comments and discussions related to these questions.

\paragraph{Task-based sample selection}
In this work, we studied the task of distinguishing M- and X-class flare producing regions from flare-quiet regions. This ``strong-vs-quiet" task focuses on the increase or the continuation of flare activity and does not require distinguishing between flares of closer energy levels like M- and C-class flares. The samples that indicate a decay in flare activity and the samples that only lead to weak flares are therefore excluded. This ensures good baseline classification performance against which our proposed predictors could be reliably compared. Our findings may not extend to other task definitions, e.g., all-clear forecasts, where flare-quiet active regions that evolve from a flare-active state are of interest \citep{barnes2016comparison,ji2020all}; and operationally-evaluated forecasts, where flare activities in the prediction period are considered as prescience and can not be used to select samples. The principal challenge to extending our analysis to these tasks is that weak- and no-flare activity annotations have higher uncertainty due to background radiation and other factors \citep{mccloskey2018flare}. The higher levels of ``label noise" when including weak flares as negative samples would make learning a reliable predictor substantially more difficult. A possible solution, and topic of future work, is to predict the continuous flare intensity level instead of the GOES flare activity class.

\paragraph{Evaluation under a realistic event rate} As mentioned in Section~\ref{sec:rus}, we rebalance the training and the validation set to prevent the predictor biasing towards the majority non-event class simply due to its volume, and we rebalance the test set to evaluate the predictor's generalization ability under the same climatological rate as it is trained. Evaluating the performance under a realistic event rate requires more work other than simply applying the predictor on a test set that is not rebalanced: predictors trained on the balanced dataset will bias towards the minority class on a test set under a realistic rare event rate, producing an undesirably high false alarm rate. One possible solution to correct such bias is to treat the class proportions as priors and apply the Bayes rule \citep{elkan2001foundations}. This method requires an accurate estimate of the true event rate of the testing period.

\paragraph{The importance and challenges of data fusion} Fusing data from multiple sources to produce more consistent, accurate, and useful information is a universal problem in astronomy. Although the astrophysics community is funding projects like DKIST \citep{rimmele2020daniel} and the Vera Rubin Observatory \citep{ivezic2019vera}, both of which will take 25--50 TB of data a day, astronomers cannot study long-term trends without including historical or old data sets (or waiting a decade for these instruments to take enough data). In this work, we took a straightforward approach to add the new data in the training set with minimal calibration, and train the models as usual. Based on our experiments, we have shown that this simple approach can result in improvement. We anticipate that, with more accurate cross-calibration between the SMARP and SHARP, the benefit of combining them may be better than demonstrated here. There are several possible ways to improve upon the fusion method:
\begin{itemize}
    \item To simulate the effect of unresolved structures in SMARP magnetograms, Gaussian blur can be applied to the higher quality SHARP magnetogram. This approach is used in comparing full-disk line-of-sight magnetograms of HMI and MDI in \citet{liu2012comparison}, in which the parameters of the Gaussian filters are tuned to minimize the root mean squared difference between them. 
    \item Point spread functions can be estimated for MDI and HMI magnetograms separately, and deconvolution can be performed to remove stray light that is instrument-specific \citep{mathew2007properties,yeo2014point}. 
    \item Magnetogram fusion can be performed in the other direction: super-resolving magnetograms in SMARP to mimic those in SHARP. Such an approach has been recently explored using deep neural networks \citep{gitiaux2019probabilistic, jungbluth2019single}. The improved overall image quality of super-resolved SMARP magnetograms could capture higher resolution magnetic field distributions and hence improve the accuracy of the active region summary parameters in SMARP. 
    \item For the active region summary parameters, we took a ``post-facto" correction approach by correcting the parameters of the same name in the two data products via linear regression. Alternatively, with fused magnetograms available, one can also re-compute the parameters on those transformed image data. This approach avoids the linear assumption and leads to parameters more consistent with the manipulated magnetograms, with the caveat that the manipulated magnetograms also suffer from the loss of information. More concretely, the effects of spatial resolution on the inferred magnetic field and derived quantities have been examined by \citet{leka2012modeling}, who found that, to preserve the underlying character of the magnetic field, post-facto binning can be employed with some confidence, albeit less so for derived quantities like vertical current density. In short, a universal and accurate fusing strategy that accounts for the instrumental spatial resolution is still hindered by our ignorance of the ground truth magnetic field structure, and the benefits and drawbacks of different fusing methods have to be evaluated case by case.
\end{itemize}

\paragraph{Machine learning with multi-source data} Learning from multi-source data is also a prevalent topic in machine learning. In our work, machine learning models are trained as usual with new data added to the training set. An alternative approach would use transfer learning: train on the additional data first, then switch to the original data for fine-tuning. In heliophysics, this idea is recently explored by \citet{covas2020transfer} in the prediction of the solar surface longitudinally averaged radial magnetic field distribution, using historical data from 1874 to 1975 in addition to newer data obtained by SoHO and SDO. 

\paragraph{Performance comparison between the LSTM and the CNN}
The active region summary parameters used by the LSTM are derived from magnetograms. In that sense, the data used by the CNN contains complete information of the data used by the LSTM. However, our experiments show that the LSTM generally has better performance. There are many potential reasons that the CNN does not perform better than, or as well as the LSTM: (1) the CNN takes in uniformly sized magnetograms whose size and aspect ratio are distorted. (2) the CNN only uses the image of the last frame in the sequence, whereas the LSTM uses all the data in the sequence; (3) the CNN learns the features by itself, wheres the LSTM uses hand-crafted parameterizations that are known to be relevant to flaring activity; (4) the CNN uses subsampled images with information loss, whereas the LSTM uses parameters derived from full resolution images; (5) the CNN has more parameters and more prone to overfitting (which reflects on the lower training loss but not validation loss of the CNN in many experiments).


\paragraph{On the comparison of flare forecast methods}
Many flare forecast studies quote the skill scores directly from other studies for comparison. Even though the forecast goal is somewhat standard and used in many studies (e.g. to predict whether there will be an M1.0+ class flare occurring in the future 24 hours), to conclude the superiority of one method against another, both methods have to be evaluated on the same dataset.
However, it is not trivial to come up with such a ``common ground" for methods to compete because research codes are not usually publicly available, and because different opinions exist on the ways the data should be processed. The difficulty in methodical comparison spawns the effort in fairly comparing existing forecasts (e.g., the ``All-Clear" workshops \citep{barnes2016comparison,leka2019comparison}) and developing common datasets (e.g., SWAN-SF benchmark dataset \citep{angryk2020multivariate}) or platforms (e.g., the FLARECAST project \citep{georgoulis2021flare}). To advocate credible comparisons, we do not quote skill scores from other studies. Instead, we follow the above studies and make our code publicly available to facilitate future comparisons.

\paragraph{On stacking ensemble}
In our experiments, stacking CNN and LSTM performs similarly to the ``select best" strategy but not significantly better in most settings. However, another stacking study  \citet{guerra2020ensemble} used a larger number of base learners and showed that most ensembles achieved a better skill score (between 5\% to 15\%) than any of the members alone. This suggests that improved performance may be obtained by training a larger number of base learners on the SMARP/SHARP data studied here.

\paragraph{Choice of the baseline in interpretability methods} Some visual attribution methods require reference input, such as Integrated Gradients and DeepLIFT. One naive choice is an image with all values equal to zero. Images of this sort imply a lack of patterns.
These are the baselines mostly used for interpretation in computer vision tasks like object detection.
In our case, the images are magnetic field component measurements, which can take on positive or negative values and a wide dynamic range, unlike normal images in real life. We choose the first image in the sequence as the reference, so that the visual attribution methods can attribute the change of prediction scores to the change of magnetic field configuration, which is of actual interest.
There are other choices of baselines. One example is input images with Gaussian noise. Using this type of reference may reveal the sensitivity of the network's prediction to local changes.
Furthermore, integration may benefit from going beyond simply linearly interpolating the reference and the input on the original image space, i.e., the 2D cartesian plane. For example, one could consider applying attribution methods to the path of time series of magnetograms. 
The Integrated Gradients calculated with this approach would integrate temporal dependency of each point-in-time in the sequence, exploiting more information about the evolution of active regions.


The authors would like to thank K. D. Leka for valuable discussions on the polarity artifacts of the line-of-sight component of the photospheric magnetic field, and on the effect of spatial resolution on magnetograms and derived quantities.
This work was supported by NASA DRIVE Science Center grant 80NSSC20K0600.

\software{Our codes for data processing, model training, and performance evaluating are openly available at Zenodo via DOI \href{https://doi.org/10.5281/zenodo.6415849}{10.5281/zenodo.6415849} \citep{sun2022zenodo} or GitHub at \url{https://github.com/ZeyuSun/flare-prediction-smarp}.}

\bibliography{main}{}
\bibliographystyle{aasjournal}


\end{document}

%% file: table_member.tex
\begin{tabular}{cccccc}
\toprule
    &  & \multicolumn{2}{c}{Group 1} & \multicolumn{2}{c}{Group 2} \\
    \cmidrule(lr){3-4} \cmidrule(lr){5-6}
    & Dataset &    \texttt{\uppercase{fused\_sharp}} &          \texttt{\uppercase{sharp\_only}} &    \texttt{\uppercase{fused\_smarp}} &          \texttt{\uppercase{smarp\_only}} \\
{} & Model &                &                &                &                \\
\midrule
\multirow{2}{*}{ACC} & CNN &  0.906+/-0.036 &  0.922+/-0.017 &  \textbf{0.901+/-0.028} &  0.877+/-0.031 \\
    & LSTM &  \textbf{0.950+/-0.012} &  0.942+/-0.016 &  \textbf{0.905+/-0.025} &  0.900+/-0.024 \\
\cmidrule{1-6}
\multirow{2}{*}{AUC} & CNN &  0.980+/-0.009 &  0.981+/-0.006 &  \textbf{0.963+/-0.017} &  0.950+/-0.020 \\
    & LSTM &  \textbf{0.990+/-0.004} &  0.986+/-0.004 &  \textbf{0.966+/-0.015} &  0.963+/-0.015 \\
\cmidrule{1-6}
\multirow{2}{*}{TSS} & CNN &  0.812+/-0.071 &  0.843+/-0.034 &  \textbf{0.802+/-0.056} &  0.754+/-0.061 \\
    & LSTM &  \textbf{0.900+/-0.023} &  0.884+/-0.032 &  \textbf{0.810+/-0.050} &  0.800+/-0.049 \\
\cmidrule{1-6}
\multirow{2}{*}{BSS} & CNN &  0.649+/-0.152 &  0.714+/-0.064 &  \textbf{0.628+/-0.114} &  0.520+/-0.121 \\
    & LSTM &  \textbf{0.799+/-0.036} &  0.775+/-0.047 &  \textbf{0.626+/-0.107} &  0.586+/-0.108 \\
\bottomrule
\end{tabular}


%% file: table_ttest.tex
\begin{tabular}{cccccc}
\toprule
    & $H_1$ & \multicolumn{2}{c}{$S_{\texttt{FUSED\_SHARP}} > S_{\texttt{SHARP\_ONLY}}$} & \multicolumn{2}{c}{$S_{\texttt{FUSED\_SMARP}} > S_{\texttt{SMARP\_ONLY}}$} \\
    \cmidrule(lr){3-4} \cmidrule(lr){5-6}
    & {} &                                              $p$-value &       $t$ &                                              $p$-value &       $t$ \\
Metric $S$ & Model &                                                        &           &                                                        &           \\
\midrule
\multirow{2}{*}{ACC} & CNN &                                               0.885787 & -1.292359 &                                               \textbf{0.001862} &  3.881137 \\
    & LSTM &                                               \textbf{0.016544} &  2.514074 &                                               \textbf{0.026797} &  2.219666 \\
\cmidrule{1-6}
\multirow{2}{*}{AUC} & CNN &                                               0.589845 & -0.233881 &                                               \textbf{0.001399} &  4.070352 \\
    & LSTM &                                               \textbf{0.000459} &  4.842485 &                                               \textbf{0.033930} &  2.074572 \\
\cmidrule{1-6}
\multirow{2}{*}{TSS} & CNN &                                               0.885787 & -1.292357 &                                               \textbf{0.001862} &  3.881135 \\
    & LSTM &                                               \textbf{0.016544} &  2.514079 &                                               \textbf{0.026796} &  2.219673 \\
\cmidrule{1-6}
\multirow{2}{*}{BSS} & CNN &                                               0.889419 & -1.314583 &                                               \textbf{0.000482} &  4.806837 \\
    & LSTM &                                               0.054812 &  1.775082 &                                               \textbf{0.000099} &  6.014784 \\
\bottomrule
\end{tabular}


%% file: table_ttest_est.tex
\begin{tabular}{ccccccccc}
\toprule
Dataset & \multicolumn{2}{c}{$\texttt{FUSED\_SHARP}$} & \multicolumn{2}{c}{$\texttt{SHARP\_ONLY}$} & \multicolumn{2}{c}{$\texttt{FUSED\_SMARP}$} & \multicolumn{2}{c}{$\texttt{SMARP\_ONLY}$} \\
\cmidrule(lr){2-3} \cmidrule(lr){4-5} \cmidrule(lr){6-7} \cmidrule(lr){8-9}
{} &               $p$-value &       $t$ &              $p$-value &       $t$ &               $p$-value &       $t$ &              $p$-value &       $t$ \\
Metric $S$ &                         &           &                        &           &                         &           &                        &           \\
\midrule
ACC        &                \textbf{0.001442} &  4.050296 &               \textbf{0.007142} &  3.028403 &                0.234866 &  0.754672 &               \textbf{0.001079} &  4.245351 \\
AUC        &                \textbf{0.003757} &  3.429557 &               \textbf{0.002527} &  3.682754 &                0.227978 &  0.779031 &               \textbf{0.000743} &  4.501005 \\
TSS        &                \textbf{0.001442} &  4.050297 &               \textbf{0.007142} &  3.028405 &                0.234865 &  0.754673 &               \textbf{0.001079} &  4.245350 \\
BSS        &                \textbf{0.005296} &  3.213872 &               \textbf{0.002645} &  3.653351 &                0.531965 & -0.082481 &               \textbf{0.005781} &  3.159315 \\
\bottomrule
\end{tabular}